\documentclass[sigplan, screen]{acmart}

\newtoggle{EXTND} 

\togglefalse{EXTND}

\toggletrue{EXTND} 

\iftoggle{EXTND}
{
\settopmatter{printccs=false,printacmref=false,printfolios=true}
\renewcommand\footnotetextcopyrightpermission[1]{}
\pagestyle{plain}
}
{
\settopmatter{}
}

\setcopyright{acmlicensed}
\acmPrice{15.00}
\acmDOI{10.1145/3314221.3314626}
\acmYear{2019}
\copyrightyear{2019}
\acmISBN{978-1-4503-6712-7/19/06}
\acmConference[PLDI '19]{Proceedings of the 40th ACM SIGPLAN Conference on Programming Language Design and Implementation}{June 22--26, 2019}{Phoenix, AZ, USA}
\acmBooktitle{Proceedings of the 40th ACM SIGPLAN Conference on Programming Language Design and Implementation (PLDI '19), June 22--26, 2019, Phoenix, AZ, USA}
\usepackage{booktabs}   
\usepackage{subcaption} 

\usepackage[]{algorithm2e}
\usepackage{listings}

\usepackage{pifont}

\newcommand{\cmark}{\ding{51}}%
\newcommand{\xmark}{\ding{55}}%

\usepackage[]{xcolor}

\usepackage{todonotes}

\InputIfFileExists{activateeditingmarks}{
}{
    \def\noeditingmarks{}
}

\definecolor{comment-red}{rgb}{0.8,0,0}
\definecolor{dark-green}{rgb}{0.0,0.4,0}
\definecolor{dark-blue}{rgb}{0.0,0.0,0.55}
\definecolor{very-dark-green}{rgb}{0.0,0.3,0}

\ifx\noeditingmarks\undefined

   \newcommand{\new}[1]{{\color{very-dark-green}{#1}}}

   \definecolor{mygrey}{rgb}{0.7,0.7,0.7}

   \newcommand{\pgwrapper}[2]{\todo[size=\scriptsize]{#1: #2}}

   \InputIfFileExists{activategreybg}{
       \definecolor{comment-red}{rgb}{0.5,0,0}
       \pagecolor{mygrey}
   }{}

  \newcommand{\rnfloat}[1]{\pgwrapper{RRN}{#1}\vspace{-4mm}}  

\else

   \newcommand{\pgwrapper}[2]{}

   \newcommand{\new}[1]{#1}

   \newcommand{\rnfloat}[1]{}  
\fi

\newcommand{\eg}{{e.g.,}}

\def\scrunchit{}
\ifx\scrunchit\undefined
  
\else

\fi

\begin{document}

\title{Sound, Fine-Grained Traversal Fusion for Heterogeneous Trees}         

\iftoggle{EXTND}{
\titlenote{Extended version of ``Sound Fine-Grained Traversal Fusion for Heterogeneous Trees,'' Sakka et al., PLDI 2019\cite{grafter}.}             
}{}

\iftoggle{EXTND}{
\subtitle{Extended version}                     
}{}


\author{Laith Sakka, Kirshanthan Sundararajah}
\affiliation{
  \department{Electrical and Computer Engineering}              
  \institution{Purdue University}            
}
\email{{lsakka, ksundar}@purdue.edu}          


\author{Ryan R. Newton}
\affiliation{
  \department{Computer Science}              
  \institution{Indiana University}            
}
\email{rrnewton@indiana.edu}          

\author{Milind Kulkarni}
\affiliation{
  \department{Electrical and Computer Engineering}              
  \institution{Purdue University}            
}
\email{milind@purdue.edu}          

\renewcommand\shortauthors{L. Sakka, K. Sundararajah, R. Newton, and M. Kulkarni}

\newcommand{\system}{{\sc Grafter}\xspace}

\newcommand*\CPP{C\kern-0.2ex\raisebox{0.4ex}{\scalebox{0.8}{+\kern-0.4ex+}}}



    \definecolor{listinggray}{gray}{0.9}
    \definecolor{lbcolor}{rgb}{0.9,0.9,0.9}

\lstdefinelanguage{grafter-lang} {
    basicstyle=\ttfamily\footnotesize,
    backgroundcolor=\color{lbcolor},
    tabsize=2,
    language=C++,
    captionpos=b,
    tabsize=3,
    numbers=none,
    frame=lines,
    breaklines=true,
    showstringspaces=false,
    keywordstyle=\color[rgb]{0,0,1},
    commentstyle=\color{red},
    stringstyle=\color{red}
}

\lstnewenvironment{grafter}
{\lstset{language=grafter-lang}}
{}

\lstnewenvironment{grafter-numbered}
{\lstset{language=grafter-lang,numbers=left,numberstyle=\tiny,numbersep=5pt}}
{}


\lstMakeShortInline[language=grafter-lang,mathescape,keepspaces,basicstyle=\ttfamily\normalsize,breakatwhitespace]@

\newcommand{\code}[1]{\lstinline[language=grafter-lang,mathescape,keepspaces,basicstyle=\ttfamily\normalsize,breakatwhitespace]{#1}}

\begin{abstract}
Applications in many domains are based on a series of traversals of tree 
structures, and {\em fusing} these traversals together to reduce the total number 
of passes over the tree is a common, important optimization technique. In 
applications such as compilers and render trees, these trees are heterogeneous: 
different nodes of the tree have different types. Unfortunately, prior work for 
fusing traversals falls short in different ways: they do not handle heterogeneity; 
they require using domain-specific languages to express an application; they rely 
on the programmer to aver that fusing traversals is safe, without any soundness 
guarantee; or they can only perform coarse-grain fusion, leading to missed fusion 
opportunities. This paper addresses these shortcomings to build a framework for 
fusing traversals of heterogeneous trees that is automatic, sound, and
fine-grained. We show across several case studies that our approach is able to allow 
programmers to write simple, intuitive traversals, and then automatically fuse 
them to substantially improve performance.

\end{abstract}



\maketitle

\section{Introduction}
\label{sec:introduction}

Many applications are built around traversals of tree structures: 
from compilers, where abstract syntax trees represent the syntactic structure a 
program and traversals of those ASTs are used to analyze and rewrite code; 
to web browsers and layout engines, where render trees express the structure 
of documents and traversals of those trees determine the location and appearance of 
elements on web pages and documents; to solving integral and differential equation
of multi-dimensional spatial functions where kd-trees are used to represent
piecewise functions and operations on those functions are implemented as tree traversals. 
There is a fundamental tension between writing these applications in the most ergonomic manner---where, for example, a compiler is 
written as dozens of individual AST-rewriting passes~\cite{nanopass,petrashko2017}---and 
writing these applications in the most performant manner---where many AST traversals must be 
{\em fused} into a single traversal to reduce the overhead of 
traversing and manipulating potentially-large programs~\cite{petrashko2017}.

In an attempt to balance these competing concerns, there has been prior work on compiler and software-engineering techniques for writing simple, fine-grained tree traversal passes that are then automatically {\em fused} into coarse-grained passes for performance reasons~\cite{sakka2017, rajbhandari2016a, rajbhandari2016b, meyerovich2010, meyerovich2013, petrashko2017}. In the world of functional programs, {\em deforestation} techniques rewrite data structure operations to avoid materializing intermediate data structures, either through syntactic rewrites~\cite{wadler1990, chin1999fusiontupling} or through the use of special combinators that promote fusion~\cite{stream-fusion}. For web browsers, render-tree passes can be expressed in high-level, attribute grammar--like languages~\cite{meyerovich2010} and then passed to a compiler that generates fused passes~\cite{meyerovich2013}. For solving differential and integral equations, computations on spatial functions can be expressed using high-level numerical operators~\cite{rajbhandari2016a} that are then fused together into combined kd-tree passes by domain-specific compilers~\cite{rajbhandari2016a, rajbhandari2016b}. In compilers, AST-rewriting passes can be restructured using special {\em miniphase} operations that are then combined (as directed by the programmer) into larger AST phases that perform multiple rewrites at once~\cite{petrashko2017}.

Previous approaches rely on programmers using special-purpose languages or programming styles to express tree traversals, limiting generality, TreeFuser, by Sakka et al., offers an alternative~\cite{sakka2017}. Programmers write {\em generic} tree traversals in an imperative language---a subset of C---with no restrictions on how trees are traversed (unlike Rajbhandari et al. who limit computations to binary trees traversed in pre- or post-order~\cite{rajbhandari2016a, rajbhandari2016b}, or Petrashko et al. who require very specific traversal structures in miniphases~\cite{petrashko2017}). TreeFuser analyzes the dependence structure of the general tree traversals to perform {\em call-specific partial fusion}, which allows {\em parts} of traversals to be fused (unlike other prior work), and integrate code motion which implicitly restructures the order of traversals (e.g., transforming a post-order traversal into a pre-order traversal) to maximize fusion opportunities. TreeFuser hence represents the most general extant fusion framework for imperative tree traversals.


Unfortunately, TreeFuser suffers from several key limitations that prevent it from 
fulfilling the goal of letting programmers write idiomatic, simple tree traversals 
while relying on a compiler to automatically generate coarse-grained efficient traversals. 
First,
TreeFuser's dependence representation requires that trees be {\em homogeneous}: 
each node in the tree must be the same data type. This means that to support trees, 
such as abstract syntax trees, that are naturally {\em heterogeneous}, 
TreeFuser requires programmers to unify all the subtypes of a class hierarchy 
into a single type---e.g., a tagged union---distinguishing between them with conditionals. 
Second, TreeFuser does not support mutual recursion---traversals written as a set of functions, rather than a single one---requiring the use of many conditionals to handle different behaviors. 
As a corollary of not supporting heterogeneous trees or mutual recursion, 
TreeFuser does not support virtual functions, a key feature that, 
among other things, allows complex traversals to be decomposed into operations on individual node types. 
These limitations require expressing traversals with unnatural code and produce spurious dependences, that can inhibit fusion.
Finally, TreeFuser does not support tree {\em topology mutation}.
While fields within nodes in a tree can be updated in TreeFuser traversals, 
the topology of the tree must be read-only. This makes it unnatural to 
express some AST rewrites (by, e.g., changing a field in a node to mark it as deleted, 
instead of simply removing the node) and impossible to express others.

\subsection{Contributions}

This paper presents a new fusion framework, \system, that addresses these limitations to support a more idiomatic style of writing tree traversals, getting closer to the goal of fusing truly general tree traversals. The specific contributions this paper makes are:

\begin{enumerate}
\item \system provides support for heterogeneous types. Rather than requiring that each node in the tree share the same type (as in prior work on fusing general traversals), \system allows recursive fields of a node in a tree to have any type, enabling the expression of complex heterogeneous tree structures such as ASTs and render trees.
\item To further support heterogeneous tree types, \system supports mutual recursion and virtual functions, allowing children of nodes to be given static types that are resolved to specific subtypes at runtime, more closely matching the natural way that traversals of heterogeneous trees are written. This support requires developing a new dependence representation that enables precise dependence tests in the presence of mutual recursion and dynamic dispatch. \system also adds support in its dependence representation for accommodating insertion and deletion of nodes in the tree.
\item \system generalizes prior work's call-specific partial fusion to incorporate {\em type-specific partial fusion}. This allows traversals to be fused for some node types but not others, yielding more opportunities for fusion. Moreover, by leveraging type-specific partial fusion and dynamic dispatch, \system's code generator produces simpler, more efficient fused traversal code than prior approaches, resulting in not just fewer passes over the traversed tree, but fewer instructions total.
\end{enumerate}


\begin{table}
\caption{Grafter in comparison to prior work. Note that \system provides finer-grained fusion than TreeFuser. We exclude syntactic rewrites of functional programs (e.g., \cite{wadler1990}), as their rewrites are not directly analogous to fusion.}  
\label{tab:summary}
\resizebox{\linewidth}{!}{

\begin{tabular}{p{3.4cm}p{1.2cm}p{1.15cm}p{1.25cm}p{1.3cm}}
{\bf Approach} & {\bf Hetero\-geneous trees} & {\bf Fine-grained fusion} & {\bf General express\-ivity} &{\bf Depen\-dence analysis} \\
\hline
Stream fusion \cite{stream-fusion} & \cmark & \xmark & \xmark &\textbf{NA}\\
Attribute grammars \cite{meyerovich2013} & \cmark & \xmark & \xmark & \cmark\\
Miniphases \cite{petrashko2017} & \cmark & \xmark & \xmark& \xmark \\
Rajbhandari et al. \cite{rajbhandari2016b} & \xmark & \xmark & \xmark&\xmark  \\
TreeFuser \cite{sakka2017} & \xmark & \cmark & \cmark & \cmark  \\
Grafter & \cmark & \cmark & \cmark & \cmark 
\end{tabular}
}
\end{table}

Table~\ref{tab:summary} summarizes Grafter's capabilities in relation to prior 
work. At a high level, prior work either supports heterogeneous trees but not 
fine-grained fusion or vice versa, while Grafter supports both, as well as the 
ability to write general traversals, analyze dependences and perform  
sound fusion automatically.

We show across several case studies that \system is able to deliver substantial performance benefits by fusing together simple, ergonomic implementations of tree traversals into complex, optimized, coarse-grained traversals.

\subsection{Outline}

The remainder of the paper is organized as follows.
Section~\ref{sec:overview} provides an overview of our fusion framework \system. 
Section~\ref{sec:design} lays out the design of \system: the types of traversals it supports, how it represents dependences given the more complex traversals, how it performs type-directed fusion, and how it synthesizes the final fused traversal(s). 
Section~\ref{sec:implementation} details the prototype 
implementation of \system. 
Section~\ref{sec:evaluation} evaluates \system across several case studies. Section~\ref{sec:related} discusses related work and Section~\ref{sec:conclusions} concludes.


\section{Grafter Overview}
\label{sec:overview}

\begin{figure*}

    \begin{subfigure}[b]{0.25\linewidth}
        \centering
        \includegraphics[scale=0.45]{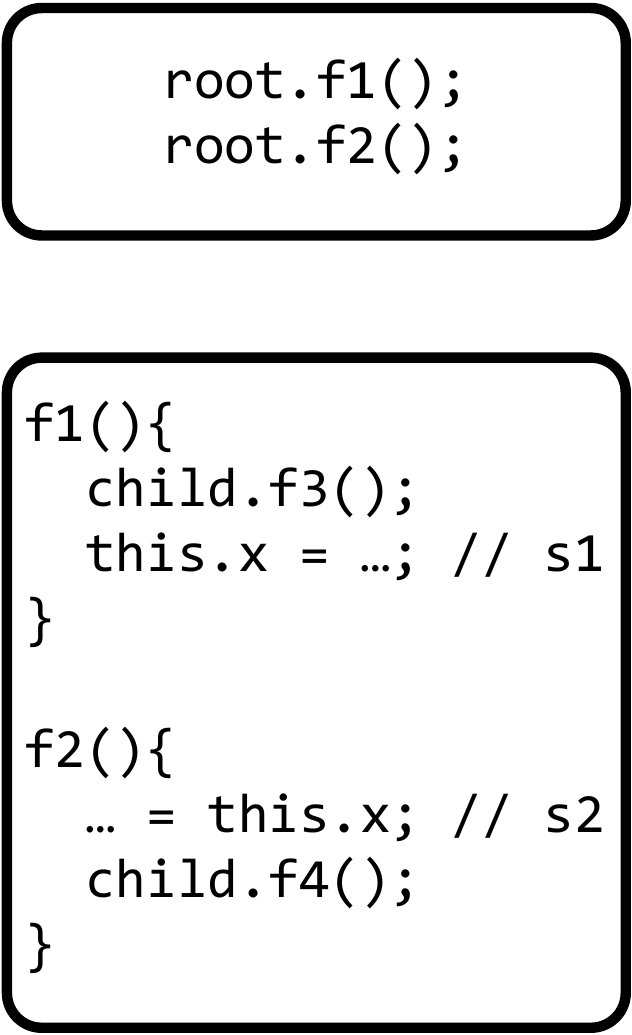}
        \caption{Unfused code}
        \label{fig:overview-a}
    \end{subfigure}%
    \begin{subfigure}[b]{0.25\linewidth}
        \centering
        \includegraphics[scale=0.45]{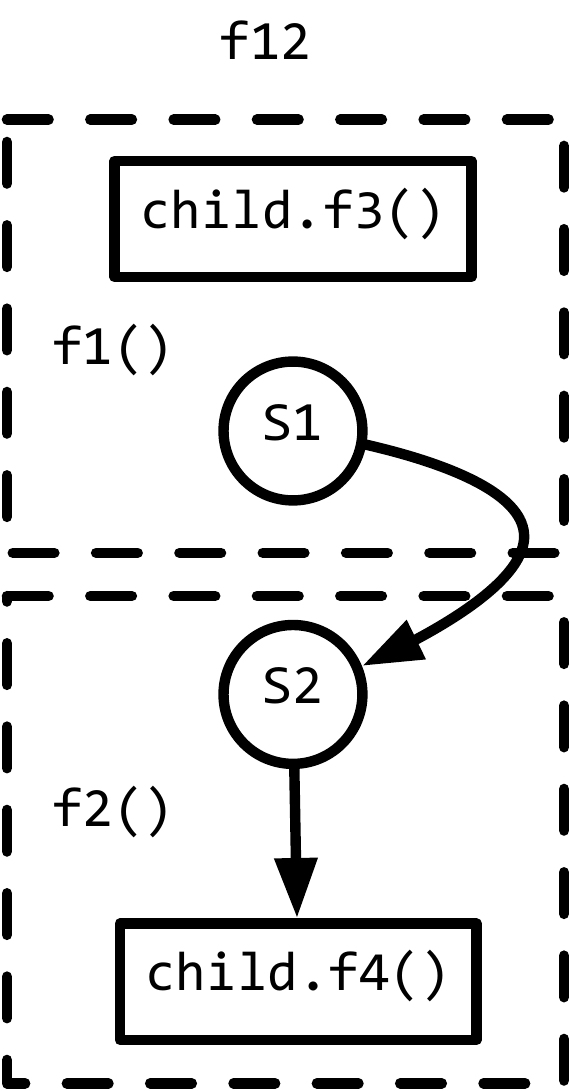}
        \caption{Outline and inline step}
        \label{fig:overview-b}
    \end{subfigure}%
    \begin{subfigure}[b]{0.25\linewidth}
        \centering
        \includegraphics[scale=0.45]{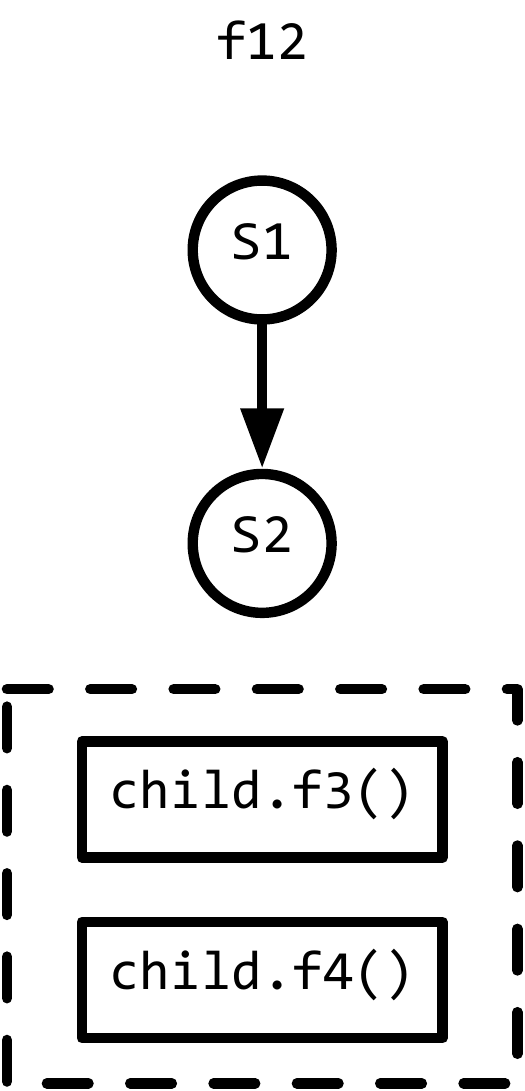}
        \caption{Reorder step}
        \label{fig:overview-c}
    \end{subfigure}%
    \begin{subfigure}[b]{0.25\linewidth}
        \centering
        \includegraphics[scale=0.45]{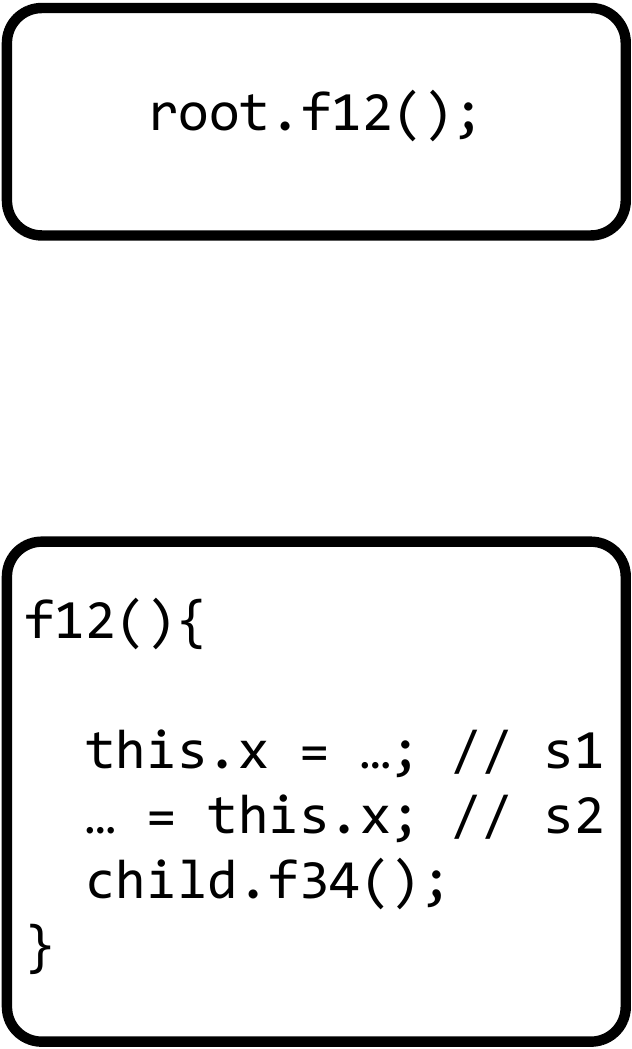}
        \caption{Fused code}
        \label{fig:overview-d}
    \end{subfigure}
    \caption{Fusing mutual traversals in \system.}

\end{figure*}
\system adopts a strategy for fusing traversals where  a programmer writes individual
 tree traversals in a standard, \CPP-like language (Section~\ref{sec:language}), as 
 functions that traverse a tree structure. The fields of each tree node can be heterogeneous, 
 and part of a complex class hierarchy, and the (mutually) recursive functions that 
 perform a tree traversal can leverage dynamic dispatch when visiting children of a node.
A full example of this style of program is shown in Figure~\ref{fig:running-example-code},
but for illustration purposes, we use a simple example shown in Figure~\ref{fig:overview-a},
with two calls to the functions $f_1$ and $f_2$. Each of those functions consists of a single statement 
($s_1$ and $s_2$, respectively), and a \textit{traversing} call on the \code{root}'s child, 
\code{child} ($f_3$ and $f_4$, respectively). These two function calls are not independent of 
each other: $s_1$ in $f_1$ updates \code{this.x} while $s_2$ in $f_2$ reads \code{this.x}.
\system performs fusion on such traversals to generate a new set of mutually-recursive functions that perform fewer traversals of the tree. 

The fusion process starts with a sequence of traversals that are invoked on the same node of a tree,
in this case \code{root}. 
Note that it is clearly safe to outline the two calls in  Figure~\ref{fig:overview-a}  into one call to 
function $f_{12}$ that executes the two functions back-to-back in their original
order as shown in Figure~\ref{fig:overview-b} (the two calls are outlined and then inlined).

\system then creates a {\em dependence graph} representation of the new function $f_{12}$. 
This dependence graph represents each call and statement of the traversals as a
vertex, and (directed) edges are placed between vertices if two
statements {\em when executing at a particular node in the tree} can access the same memory location
(\system borrows this representation from TreeFuser~\cite{sakka2017}.)
In Figure~\ref{fig:overview-b}, there is a dependence between $s_1$ and $s_2$. We also assume,
for illustration purposes, that there is a dependence between $s_2$ and the call $child.f_4()$.  
\system finds dependences between calls and statements by considering the transitive closure of what calls 
may access. Section~\ref{sec:design-dependences} describes how \system analyzes accesses to 
find dependences and construct the dependence graph.

$f_{12}$ is now a new, single function that performs multiple pieces of work on \code{root},
and invokes multiple traversal functions on \code{root.child}.
To optimize the body of $f_{12}$  its desirable to have $s1$, $s2$ executed closer to each other for
locality benefits---if they access shared fields of \code{root}, then that data is likely to remain in 
cache. Furthermore, if the calls $f_3$ and $f_4$ on \code{child} are back-to-back then we can further 
fuse them into one call and both save a function invocation and ``visit'' \code{root.child} only once,
 instead of twice. 

The key challenge to performing this fusion is that the reordering necessary to
bring statements closer together and, more importantly, to bring traversal calls closer 
together, is not always safe. \system performs reordering for the statements using 
the dependence graph, trying to bring traversals of the same child closer to each other
without reordering any dependence edges (and hence violating dependences). 
This reordering is done by grouping the traversal calls that visit the same node together.
Figure~\ref{fig:overview-c} shows the results of  such re-ordering.

Note that in this example the  reordering step changes the traversal $f_1$ from a post-order 
traversal to a pre-order traversal ($s1$ is executed at parents before their
children in $f_{12}$, while it is executed at the children before their parents in the original function $f1$).
In other words this reordering involves performing implicit code motion that can safely
changes the schedule of the traversals for the purpose of achieving more fusion. 

As calls are grouped together, \system is presented with {\em new} sequences of 
functions that this same fusion process can be applied to. In our example, the two 
calls grouped in the dashed box in Figure ~\ref{fig:overview-c} are invoked on the same 
node of the tree (\code{root.child}), so \system can {\em repeat} this merging process,
creating a new merged function $f_{34}$, building a new dependence graph for the merged function, 
rearranging its statements, and so on. Each time this re-ordering is performed, more and more operations 
from multiple logical traversals on the same node(s) of the tree are brought closer
together, improving locality, and more and more function invocations from multiple logical
traversals are collapsed, reducing invocation overhead and the total number of times the collection of traversals visit nodes of the tree.

 If \system encounters a sequence of functions that has been fused before, of course, it can simply 
 call the already-fused implementation. \system bounds the amount of functions that 
 can be fused together to ensure this process terminates. Section~\ref{sec:design-fusion} 
 describes in detail how \system performs its fusion, including how it handles virtual functions.





Note that encountering cases where an already created fused function is being called again is the key for having significant performance improvement, since that means that the locality and overhead enhancements will be achieved recursively. In the limit, instead of 2 traversals visiting each node of the tree once each, we will have a single traversal that visits each node only once---{\em total fusion}. But any amount of collapsing  still promotes locality and reduces node visits.

The end result of \system's fusion process is a set of mutually recursive functions that together form a partially-fused traversal. Crucially, these fused functions are analyzed on a per-type basis, meaning that fusion can occur {\em partially}---not all sequences of calls in a function need to be fused---and {\em type-specifically}---fusion can occur for some concrete instantiations of virtual functions, but not others. This process leads to more fine-grained, precise fusion decisions than prior work such as TreeFuser.

The following section describes the details of \system fusion process in details.

\section{Design} 
\label{sec:design}

This section describes the design of \system's components in detail. In
particular, we first describe how \system analyzes traversal functions to
identify dependences, allowing it to build the dependence graph representation
used to drive fusion (Section~\ref{sec:design-dependences}). Then we explain how
\system uses the dependence representation to synthesize new, fused functions
(Section~\ref{sec:design-fusion}).
But first, we explain the language that \system uses to express its traversals.

\begin{figure}
\begin{grafter-numbered}
  int CHAR_WIDTH; 

  class Element {
    Element *Next;
    int Height = 0, Width = 0;
    int MaxHeight = 0, TotalWidth = 0;
    virtual void computeWidth(){};
    virtual void computeHeight(){};
  };
  
  class TextBox: public Element {
    String Text; 
    void computeWidth(){
      Next->computeWidth(); 
      Width = Text.Length;
      TotalWidth = Next->Width + Width;
    };
    void computeHeight(){
      Next->computeHeight(); 
      Height = Text.Length * (Width/CHAR_WIDTH) + 1;
      MaxHeight = Height;
      if(Next->Height > Height)
        MaxHeight = Next->Height;
    };
  };
  
  class Group: public Element {
    Element *Content;
    BorderInfo Border; 
    void computeWidth(){
      Content->computeWidth(); 
      Next->computeWidth(); 
      Width = Content->Width + Border.Size*2; 
      TotalWidth = Width + Next->Width;
    };
    void computeHeight(){
      Content->computeHeight(); 
      Next->computeHeight(); 
      Height = Content->MaxHeigh + Border.Size*2
      MaxHeight = Height;
      if(Next->Height > Height )
        MaxHeight = Next->Heigh;
    };
  };
  
  class End: public Element {
  };
  
  int main(){
    Element *ElementsList = ...; 
    ElementsList->computeWidth();
    ElementsList->computeHeight();
  }
\end{grafter-numbered}
\vspace{-1em}
\caption{An example of a program written in \system.}
\vspace{-1em}
\label{fig:running-example-code}
\end{figure} 
  

\subsection{Language}
\label{sec:language}

The language programmers use to write traversals in \system is a subset of \CPP{}, allowing programmers to integrate fusible tree traversals in larger projects.

A tree in \system is defined as an annotated \CPP{} class, where instances of the class represent tree nodes. We call any such annotated class a {\em tree type}. Figure \ref{fig:lang-types} shows the grammar for defining tree types. Children of a tree node are pointers to other tree types (not necessarily the same type as, or a subtype of, the node itself). Tree nodes can also store other (non-child) objects and primitive fields---we call these {\em data fields}.

\system traversals are written as member methods of tree types, implicitly ``visiting'' the tree node they are invoked on, and calling other traversal functions to continue the traversal. We call these {\em traversal methods} \footnote{For completeness, \system allows other methods to be defined for tree types, but if they are not explicitly annotated as traversal methods, \system will not consider them for fusion.}.
In order to support tree children with abstract types,
a tree in \system can inherit fields and virtual traversal methods from other tree types and can specialize inherited virtual traversals by overriding them.

Figure \ref {fig:traversal-def} shows the grammar for defining a traversal method in \system. Parameters of traversals are objects or primitives and are passed by value;
furthermore, \new{without loss of generality, we assume}
traversals do not have a return value\footnote{This restriction on return values simplifies \system's design, but is mostly an implementation detail.}. The body of the traversal is a sequence of, non-traversing statements ({\em simple} statements) interleaved with traversing statements ({\em traverse} statements), which are function calls to traversal methods invoked on the traversed node or one of it's children.

Assignment in \system only allows writing to data fields, and hence tree nodes can not be modified in an assignment statement. Local variables in the body of the traversal can either be data definitions (primitive or objects), or aliases to tree nodes (rules \ref{eq:local}, \ref{eq:alias}). Note that an alias variable is a constant pointer to a tree node that can only be assigned once to a descendant tree node and cannot be changed. These local variables make it easier to write traversals while precluding the need for a complex alias analysis\footnote{\system could allow more general assignment statements, coupled with a sophisticated alias analysis, but such support is orthogonal to the goals of this paper, so we do not provide it.}.

\begin{figure}
    \small
        \begin{flalign*}
          &s \in \text{<data-ref>} ::=s_1\ |\ s_2\ |\ \dots&t \in \text{<tree-ref>}\ \ ::=t_1\ |\ t_2\ |\ \dots&\\
          &c \in \text{<child-ref>}  ::=c_1\ |\ c_2\ |\ \dots&f \in \text{<traversal-ref>}:=f_1\ |\ f_2\ |\ \dots&\\
          &l \in \text{<alias-ref>}:=l_1\ |\ l_2\ |\ \dots&p \in \text{<pure-func>}:=p_1\ |\ p_2\ |\ \dots&
    \end{flalign*}
    \vspace{-0.5cm}
    \begin{subfigure}[a]{\linewidth}

    \begin{flalign}
        &\text{<prim>}\   \ \ \  ::= \ \text{\textbf{int} | \textbf{float} | \textbf{bool} | \textbf{double} | \textbf{char}}&\\
        &\text{<data-def>} ::=\ (\ \text{<prim>}\ |\ \text{<c++-class-ref>}) \ \ s &\\       
        &\begin{aligned}\label{eq:tree-def}
        \text{<tree-def>\ \ }  ::=&\textbf{\ \_tree\_ class\ } \ t \ [ \boldsymbol: \textbf{public}\ t(,\ \textbf{public}\ t)\mbox{*}]\boldsymbol{\{}\\  
                                 & (\ \textbf{\_traversal\_}\ \text{<traversal>}\\
                                 & |\ \textbf{\_child\_}\ t\  \boldsymbol{*} \ c\ ;|\ \text{<data-def>})\mbox{*} \textbf\} \textbf;
       \end{aligned}&
       \end{flalign}

       \caption{Types definition in \system.}  
       \label{fig:lang-types}
            \vspace{1em}

    \end{subfigure}

    \begin{subfigure}[b]{\linewidth}
        \small

        \begin{flalign}
        &\begin{aligned}
        \text{<traversal>}\ \ \ \ \ :=&\ [\textbf{virtual}]\ \textbf{void}\ \ f\ \textbf(\\
                                 &\ [\text{<data-def>}(,\text{<data-def>)}\mbox{*}] \textbf{)\ \{} \\
                                 &\ \text{<stmt>}\mbox{+}\ \textbf{\}}\\
        \end{aligned}&\\
        &\text{<stmt>}\ \ \ \ \ \ \ :=\ \text{<traverse-stmt>}\ |\ \text{<simple-stmt>}&\\
        &\begin{aligned}
            \text{<simple-stmt>} :=&\  \text{<if-stmt>}\    |\ \text{<delete-stmt>}\ | \text{<new-stmt>}\\
                                   & |\ \text{<assignment>}\ |\ \text{<local-def>}\  |\ \text{<alias-def>}\\
                                   &|\ \text{<pure-call>  } |\ \textbf{return ;}
       \end{aligned}&\\
       &\begin{aligned}\label{eq:recStmt}       
        \text{<traverse-stmt>}\ \ \ \ :=\ &\textbf{this[} \textbf{-->}c\textbf{]-->}f\textbf{(}
        \text{[<expr>(,\ <expr>)\mbox{*}]}\textbf{)}\
        \end{aligned}&\\\label{eq:del}
        &\text{<delete-stmt>} :=\ \textbf{delete} \ \text{<tree-node>}\ \boldsymbol{;}\\
       &\begin{aligned}\label{eq:new} 
        \text{<new-stmt>} \ \  :=&\ \text{<tree-node>}\ \boldsymbol{=}\textbf{ new}\ t \textbf{()}\ \boldsymbol{;}
        \end{aligned}&\\
        &\text{<assignment>} ::=\ \text{<data-access>}\ \boldsymbol{=}\ \text{<expr>}\  \textbf{;}&\\
        &\text{<pure-call>}  ::=\  p\ \textbf{(} \text{[<expr>(,\ <expr>)\mbox{*}]}\textbf{);}\ \label{eq:pure-call} &\\
        &\begin{aligned}
            \text{<if-stmt>}\ \ \ \ \ \ \ :=&\ \text{ \textbf{if (}<expr>\textbf{) then \{ }<simple-stmt>\mbox{*} \textbf{\}} }\\ 
                                &\textbf{\ else \{}\ \text{<simple-stmt>\mbox{*}}\ \textbf{\}}
        \end{aligned}&\\
        &\text{<local-def>} \ ::=\ \text{<data-def>}  \boldsymbol{;}\label{eq:local}\\
        &\begin{aligned}\label{eq:alias}
           \text{<alias-def>} ::=&\ t \boldsymbol{*} \textbf{const}\ l \boldsymbol{=} \text{<tree-node> } \boldsymbol{;}\\
        \end{aligned}&\\
        &\begin{aligned}\label{eq:pure-def}
            \text{<pure-func>}\  :=&\  (\ \text{<prim>}\ |\ \text{<c++-class-ref>}) \ p\ \textbf{(}\\
                                  & [\text{<data-def>}(,\text{<data-def>)}\mbox{*}] \textbf{)}\ \{  \emph{c++ code.}\ \}\\
        \end{aligned}&
    \end{flalign}
     \vspace{1em}

    \caption{Traversals definition and statements in \system.}  
    \label{fig:traversal-def}       
\end{subfigure}

\begin{subfigure}[c]{\linewidth}
    \small

    \begin{flalign}
        &\begin{aligned}
        \text{<expr>} :=&\ \text{<const>} \ |\ \text{<data-access>}\ |\ \text{<bin-expr>} |\ \text{<pure-call>}
        \end{aligned}&\\
        &\text{<tree-node>}\ \ \ :=(\textbf{this}\ |\ l\ |\ \text{<cast-expr>})(\textbf{-->}c)\mbox{+}&\\        
        &\text{<data-access>} :=\ \text{<on-tree>}\ |\ \text{<off-tree>}&\\
        &\text{<off-tree>}\ \ \ \ \ \ := s(\boldsymbol{.}s)\mbox{*}&\\
        &\text{<on-tree>}\ \ \ \ \ \  :=(\textbf{this}\ |\ l\ |\ \text{<cast-expr>})(\textbf{-->}c)\mbox{*}\text{(} \boldsymbol{.}s)\mbox{+}&\\
        &\begin{aligned}
        \text{<cast-expr>}\ :=\ &\textbf{static\_cast}\boldsymbol{<}t\boldsymbol{*>}\text{\textbf{(}<tree-node>\textbf{)}}&
        \end{aligned}&\\
        &\begin{aligned}
        \text{<bin-expr>}\ \  :=&\ \text{<expr>}\ \text{<c++-binary-op>}\ \text{<expr>}\\ 
        \end{aligned}&
        \end{flalign}

    \caption{Expressions in \system.}     
    \label{fig:exp-def}   
\end{subfigure}
\vspace{-1em}
\caption{Language of \system.}     
\vspace{-1em}
\end{figure}

\system uses \code{new} and \code{delete} \CPP{} language constructs to support leaf mutations (constructs \ref{eq:del} and \ref{eq:new} in Figure~\ref{fig:traversal-def}). The \code{new} statement allows a new tree node to be constructed and assigned to a specific child field of the current tree node, and the standard \CPP \code{delete} statement is permitted for deleting child fields (subtrees).
\system accepts such statements only if the the trivial constructor or destructor is called, i.e., user-defined constructors or destructors are not permitted.

\system allows traversals to invoke pure functions; those functions can have an object or primitive return values, and accepts object and variables as parameters. The bodies of those functions are not analyzed, and the pure annotation indicates to \system that those functions can be considered as read-only functions. Equations~\ref{eq:pure-def}, and~\ref{eq:pure-call} shows the usage of such functions.


The key operation performed by \system traversal methods is {\em accessing data and child fields}. Reads and writes to these fields, whether directly at a node or through a chain of pointer dereferences.


Accessing a variable in \system is done through an \emph{access path}. An access path is a sequence of member accesses starting from a field or a node. Accesses paths can be classified into {\em tree node} and {\em data accesses} (see Figure~\ref{fig:exp-def}). Data accesses can be further classified, based on the location of the accessed variable, into \textit{on-tree} and \textit{off-tree} accesses. The former are accesses that start at member fields of the current node (and hence are parameterized on \code{this}, the node the function is called on) or local variables in the current function, while the latter are to global data (meaning that all invocations of this function will access the same location, regardless of which node the function is executing at).
%
Any local alias variables can be recursively inlined in the access path until the access path is only a sequence of member accesses.
Access paths can also be classified to reads and writes in the obvious way. Note that <tree-node> accesses appear as writes only in the new and delete statements.

Figure \ref{fig:running-example-code} shows an example of a program written in \system (we elide the annotations for brevity). This program consists of a tree of \code{Element}s that can be \code{TextBox}es or \code{Group}s of \code{TextBox}es (with a special sentinel \code{End} type representing the end of a chain of siblings). 
Each \code{Element} can point to a sibling \code{Element}, and a \code{Group} element can contain content elements. All elements have heights and widths, that are computed by the traversals \code{computeHeight} and \code{computeWidth}, respectively.

\subsection{Dependence Graphs and Access Representations}
\label{sec:design-dependences}

The primary representation that \system uses to drive its fusion process is the dependence graph~\cite{sakka2017}. As described in Section~\ref{sec:overview}, this graph has one vertex for each {\em top level} statement\footnote{In other words, one vertex for each \code{<stmt>} construct, as shown in Figure~\ref{fig:traversal-def}. Note that this means each function is a simple sequence of top-level statements.} and edges between statements if there are dependences between them. More precisely, an edge exists between two vertices $v_1$ and $v_2$, arising from functions $f_a$ and $f_b$ ($f_a$ and $f_b$ could be the same function) if, when invoking $f_a$ and $f_b$ {\em on the same tree node} (i.e., when \code{this} is bound to the same object in both functions), either:
\begin{enumerate}
\item $v_1$ and $v_2$ may access the same memory location with one of them being a write; or
\item $v_1$ is control dependent on $v_2$ (in \system's language, this can only happen if $v_1$ and $v_2$ are in the same function and either $v_1$ or $v_2$ could \code{return} from the function).
\end{enumerate}

So how does \system compute these data dependences?

\subsubsection{Access automata}
To compute dependences between statements in different traversals, the first step is for \system to capture the set of accesses made by any statement or call in a given traversal function. To do so, \system builds {\em access automata} for each statement. These can be thought of as an extension of the regular expression--based access paths used by prior work~\cite{sakka2017,weijiang15} to account for the complexities of virtual function calls and mutual recursion.

An access path for a simple statement such as \code{n.x = n.l.y + 1} is straightforward. The statement {\em reads} $n.l.y$ and {\em writes} $n.x$. A simple abstract interpretation suffices to compute these access paths (intuitively, we perform an alias analysis on the function using access paths as our location abstraction~\cite{larus88,wiedermann07}). The abstract interpretation associates with each local variable an access path, or set of access paths when merging across conditionals, aliased to that variable. At each read (or write) of a variable, the access path(s) are added to the read (or write) set of access paths for that statement. \system computes the set of access paths for each top level simple statement in each traversal function. We do not elaborate further on this process, as this analysis is standard (and is similar to TreeFuser~\cite{sakka2017}).

The more complicated question is how to deal with building access paths for traversing function calls.
Our goal is to build a representation that captures {\em all possible access paths} that could arise as a result of invoking the function.
%
Rather than trying to construct path expressions to summarize the behavior of function calls, \system directly constructs {\em access automata} to account for this complexity. Note that these access automata are not quite like the aliasing structures computed by Larus and Hilfinger~\cite{larus88}, because \system's representation is deliberately parameterized on the current node that a function is invoked on.
We describe how \system builds these automata next.


\subsubsection*{Building access automata for statements}

Each top level statement in \system has six automata associated with it that represent reads and writes of local, global and tree accesses (including <on-tree> and <tree-node> accesses) that can happen during the execution of the statement. 

\system starts by creating primitive automata. For each access path, a primitive automaton is constructed which is a simple sequence of states and transitions. Transitions in the automata are the member accesses in the primitive access path except for two special transitions: (1) the \emph{traversed-node} transition which appears only at the start of an on-tree access and replaces \textbf{this}, and (2) the ``\emph{any}'' transition that happens on any member access.

If a primitive access path is a read, then each prefix of the primitive access is also being read, and accordingly, each state in the primitive automata is an accept state except the initial state. If a primitive access is being written, then only the full sequence is written to while the prefixes are read.

There are some special cases to deal with while constructing primitive automata. If an access is off-tree and ends with a non-primitive type (a \CPP{} object), then accessing that location involves accessing any possible member within that structure.  Such cases are handled by extending the last state with a transition to itself on any possible member using an \emph{any} transition. Likewise, tree locations that are manipulated using \emph{delete} and \emph{new} statements, writes to any possible sub-field accessed within the manipulated node and their automata uses \emph{any} transition to capture that.

 \begin{figure}
    \centering
     \includegraphics[scale=0.45]{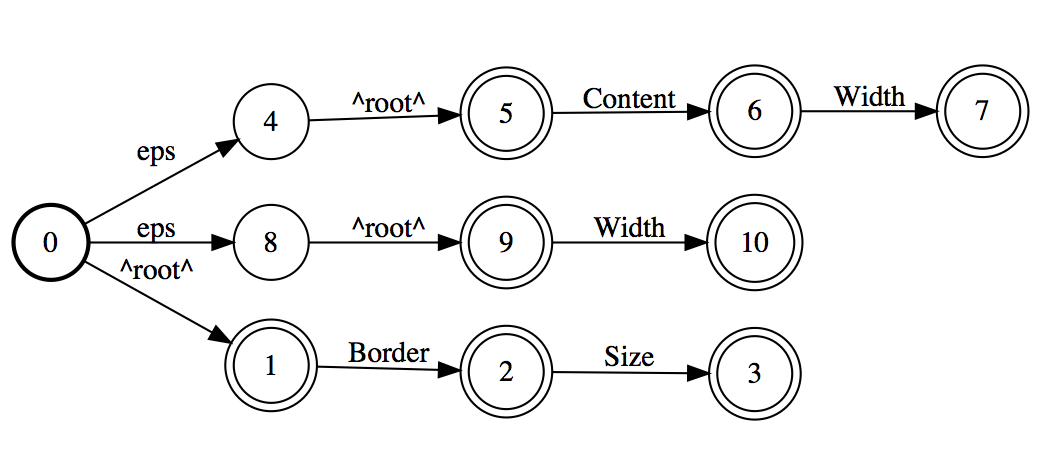}
     \vspace{-1.7em}
     \caption{Automata that represents summary of read accesses for a simple statements. \emph{eps} is epsilon transition, and \emph{root} is the traversed node transition.}
     \vspace{-1em}
     \label{fig:simple_stmt_automata}
\end{figure}

After the construction of the primitive automata, access automata of simple statements can be constructed from the union of the primitive automata. For example, the tree reads automaton for a simple statement is the union of the primitive automata of the tree read accesses in the statement. Figure \ref{fig:simple_stmt_automata} shows the tree read automaton for the statement: 
\begin{verbatim}
    Width = Content->Width + Border.Size*2; 
\end{verbatim}

\paragraph{Finding dependences between statements} These  automata provide the information needed to find dependences between statements. Because each statement's automata captures the full set of access paths read (or written) for a statement, and we are interested in whether the statements have a dependence {\em when invoked on the same tree node}, we can simply intersect the write automaton for a statement with the read and write automata for another statement to determine if a dependence could exist---a non-empty automaton means the two statements could access the same location.

\subsubsection*{Building access automata for traversing calls}
Representing accesses of traversal calls is not as simple. For a given call statement we want to construct a finite automaton that captures {\em any possible access path that could arise during the call relative to the tree node being traversed by caller}---including the fact that a call may invoke more traversals.

When building the access automaton for a traversal call, \system first creates a call graph that includes all the possibly (transitively) reachable functions from that call. We first note that any {\em off-tree} data accesses made by any of these reachable functions are, inherently, not parameterized by the receiver of the traversal calls---regardless of when and where the function gets called, those access paths will be the same. Thus, we can simply union those automata together for the functions in the call graph to capture those accesses.

The situation is more complicated for {\em on-tree} accesses, as those are parameterized by the receiver of the call, \code{this} (i.e., the node that is being traversed).
To find the accessed locations relative to \code{this}, we  need to find two things: the functions that are reachable during the call (to know which statements are executed), and the tree nodes that  those functions are invoked on, relative to \code{this}. 

Consider building the access paths for some function $f$. For each function $q$ reachable from $f$, the sequence of children traversed to get to the invocation of $q$ gives an access path for the node that $q$ is invoked on, relative to the receiver of $f$. This access path can be pre-pended to the statements access paths in $q$ to produce the access paths relative to \code{this} (the traversed node in $f$). For example,
when a function $f$ invokes another function $g$ on a child $x$ of \code{this}, it invokes \code{x.g()}; the receiver object of $g$ is \code{this.x}. Thus, to incorporate the effects of $g$ in to the access paths of $f$, we can prefix the access paths of $g$ with \code{this.x}. If $g$ in turn calls $h$ through child $y$, then we can prefix the access paths of $h$ with \code{this.x.y} and add them to the access paths of $f$.

\begin{figure}
    \begin{subfigure}[a]{\linewidth}
        \centering
         \includegraphics[scale=0.5]{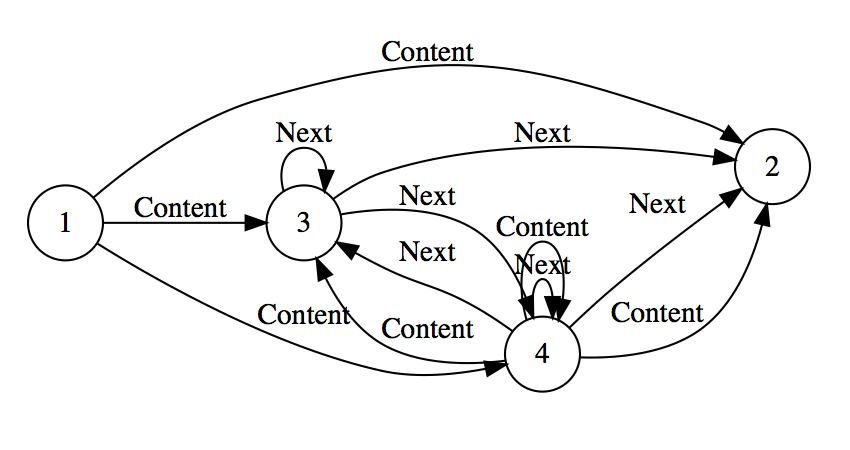}

         \vspace{-1.7em}
         \caption{Step 1: Create labeled call graph. States 2, 3 and 4 corresponds to computeWidth() functions for End, TextBox and Group types respectively.}
         \label{fig:call-writes-step1}
     \end{subfigure} %
     \begin{subfigure}[b]{\linewidth}
        \centering
         \includegraphics[scale=0.4]{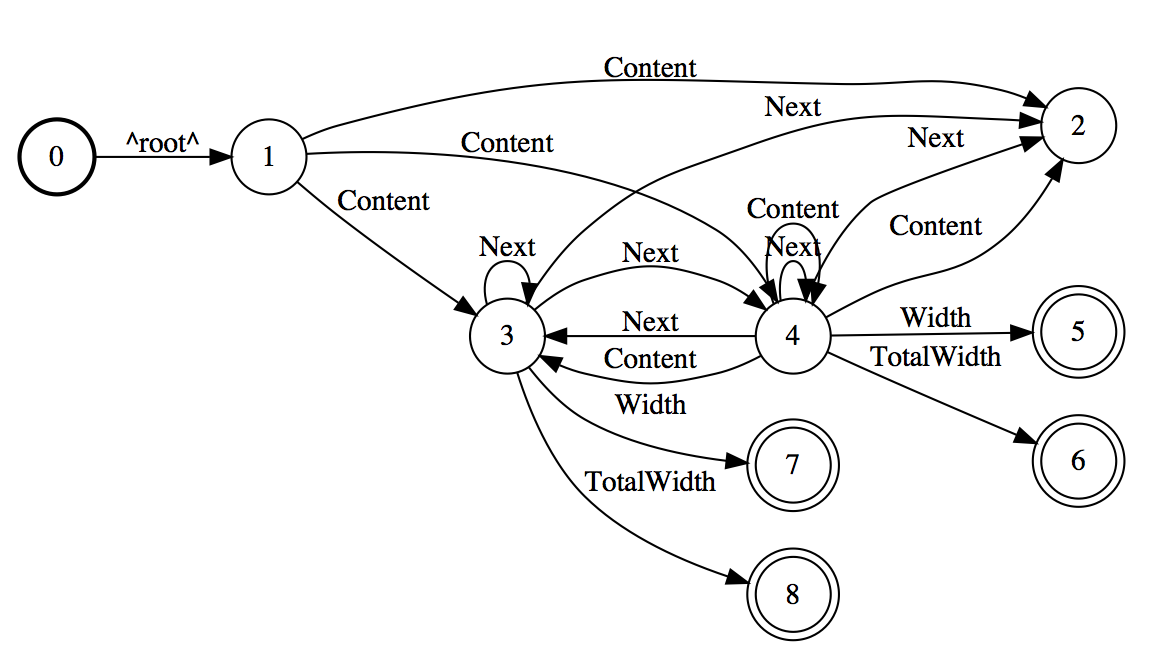}
         \vspace{-1em}
         \caption{Step 2: Attach simple statements' automata and traversed node transition.}
         \label{fig:call-writes-unsimple}
     \end{subfigure} %
     
     \begin{subfigure}[c]{\linewidth}
        \centering
        \includegraphics[scale=0.5]{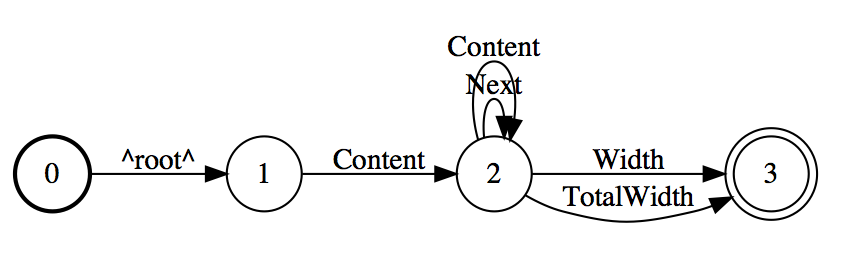}
        \vspace{-1em}
        \caption{Step 3: Minimize automata.}
        \label{fig:call-writes-simple}
    \end{subfigure} %
    \vspace{-1em}
    \caption{Construction of tree writes automata for the traversing statement
    \emph{content->computeWidth()}. \emph{root} is the traversed node transition.}
    \vspace{-1em}
    \label{fig:call-writes}
\end{figure}

To account for on-tree accesses, \system takes the call graph for the traversal call, and labels each edge with the traversed field that is the receiver for that call. Figure~\ref{fig:call-writes-step1} shows the call graph that is generated for @Content->computeWidth()@ from our running example. Paths in this graph thus correspond to possible sequences of child-node traversals to reach each function in the graph. For each function, \system then attaches all of the {\em statement} automata (see the previous section) for that function to the corresponding node in the call graph. This has the effect of treating the regular language from the statement automata as the suffix attached to the prefix that designates the receiver object. Figure~\ref{fig:call-writes-unsimple} shows the resulting automaton, and Figure~\ref{fig:call-writes-simple} shows the reduced version.

The pseudo code in Figure~\ref{fig:ontree-automata-psudo} illustrates  the construction process in detail starting from a traversing statement. First, the algorithm adds the \emph{traversed-node} transition. After that, for each possible called function that corresponds to each possible dynamic type of the called child, a state that represents all the accesses with in that function is created, and a transition on the traversed child is added. Accesses within a function are the union of the accesses of all the statements in the body of the function. Hence, for the function's traversing statements, the process will be called recursively. To guarantee the termination of such process, accesses of a unique function do not need to have more than one corresponding state in the automata. If a function is already in the automaton, then a transition to the existing state is added. Since there is only a finite number of function definitions the process is guaranteed to terminate.

\SetAlFnt{\footnotesize}
\SetAlCapFnt{\footnotesize}
\SetAlCapNameFnt{\footnotesize}

\begin{algorithm}[t]
    \DontPrintSemicolon
    \SetKwFunction{FisCallStmt}
    {isCallStmt}
    \SetKwFunction{FBuildExtended}
    {buildExtendedOnTreeAutomata}
    \SetKwFunction{Ffound}
    {found}
    \SetKwFunction{FBuildExtended}
    {buildExtendedOnTreeAutomata}
    \SetKwFunction{FgetActualCalledFunction}
    {getActualCalledFunction}
    \SetKwFunction{FaddRootNodeTransition}
    {addRootNodeTransition}
    \SetKwFunction{FAppendCall}{appendCall}
    \SetKwFunction{FappendSimpleStmt}{appendSimpleStmt}
    \SetKwFunction{FappendStmtAccesses}{appendStmtAccesses}

    \SetKwProg{Fn}{Function}{}{}
    \SetKw{KFor}{For}{}

    \Fn{\FBuildExtended(CallStmt) :}{
        Automata.addRootNodeTransition(0, 1)\;
        \FAppendCall{CallStmt, 1}\;
    } 
\;

    \Fn{\FAppendCall(CallStmt, State) :}{
        \For{Arg : CallStmt.Arguments() }{
            appendExprAccesses(Arg, 0);
        }
        \For{T : CallStmt.VisitedNode.PossibleTypes()}{
            F = getActualCalledFunction(T, CallStmt)\;
            \If{! FunctionToState.found(F)}
            {
                FunctionToState[F] = NewState = Automata.addState()\;
                \For{Stmt : F.body() }{
                    \uIf{\FisCallStmt{Stmt}}{
                        appendCall(Stmt, NewState);
                    }\Else{
                        appendStmtAccesses(Stmt, NewState);
                    }
                }
            }
            
            Automata.addTransition(CallStmt.VisitedNode,  FunctionToState[F]);
        }
    } 
    \caption{Construction of tree access automata for traversing statements.}
    \label{fig:ontree-automata-psudo}
\end{algorithm}

%

Note that the constructed automata  handle the possibility of non-statically-bounded trees; whenever we encounter a function we already have created a state for, we add a "back edge" in the automaton to the state that corresponds to that function. Unbounded recursion is hence represented by loops in the automaton.

\paragraph{Finding dependences between statements and calls} The access automata constructed for calls are no different than those constructed for statements. By using the call graph to construct access paths for receiver objects of functions, all the access paths generated by the final access automata are rooted at the same receiver object as the automata for statements. Hence, finding dependences between statements and calls (or calls and calls) can be achieved by intersecting the automata and testing for emptiness.

\subsection{Fusing Traversals}
\label{sec:design-fusion}


\paragraph{Overview}

At a high level, \system performs fusion by repeatedly invoking the following steps:

\begin{enumerate}
\item Find a sequence $L$ of consecutive traversal functions invoked on the same tree node $n$.
\item {\em Outline} these traversal functions into a new function $f_L$ that is called on $n$. If such an $f_L$ has already been created in an earlier iteration of this process, simply call the existing $f_L$.
\item {\em Inline} each of the individual traversal functions in $f_L$ to expose the work done on the node $n$.
\item {\em Reorder} the statements in $f_L$ to bring statements that access the same fields closer together {\em and} to create new sequences of traversal functions invoked on the same tree node (typically, some child node of $n$).
\item Repeat the process for these newly created sequences of calls.
\end{enumerate}

These steps constitute the fusion algorithm of \system. In particular, every time a sequence $L$ is encountered {\em more than once}, and hence an existing $f_L$ can be reused, \system has exploited an opportunity for fusion. We now explain this fusion process in more detail, and also sketch a proof of correctness.

\paragraph{Details}
Fusion starts with a sequence of traversal functions that are invoked at the same tree node (e.g., the @root@). \system searches for such candidates in the compiled program and initiates the fusion process for each of them. For example, the sequence \code{ElementsList->computeWidth();} followed by \code{ElementsList->computeHeight()} in Figure~\ref{fig:running-example-code} line 51.

Because a given function may be virtual, \system first computes all possible sequences of {\em concrete} functions that may be invoked as a result of a sequence of function calls. For each type $T$ that the sequence of calls can be invoked on, \system constructs a sequence $L$ of concrete calls. In our example, there are three, depending on whether @ElementList@ points to a @TextBox@, @Group@ or @End@.

For each function  sequence  $L$ a fused function with label $f_L$ is created (if one has not already been generated).
If the label $f_L$ does not already exist, then its corresponding function needs to be generated. A  dependence graph $G_L$ is constructed for the statements in the traversals in $L$. In other words, the fused function is essentially a function containing the {\em inlined} statements from each call in the sequence $L$, in order. Note that a sequence $L$ may contain the same static function more than one time (i.e., the same function can be invoked on a given node of a tree more than once). In this case, $G_L$ contains statements from multiple copies of that function, and the statements from the two copies are treated as coming from different traversal functions.

Once $G_L$ is constructed, the statements (nodes) in the statement can be reordered as long as no dependences are violated (as long as a pair of dependent statements are not reordered). \system thus reorders the statements to try and group invocations on the same node together. It then generates the fused function code, as explained in the next section. This newly generated function has grouped traversals invocations on the same node together (and these invocations may have come from different functions than the original sequence $L$), creating new sequences of functions. \system then process these new sequences of functions to generate more fused functions. Whenever \system encounters a sequence of functions it has seen before, it does not need to generate a new function, but instead inserts a call to the already generated function. Crucially, if this new sequence is {\em the same as for a function currently being generated}, \system just inserts a recursive call to that function.

The end result is a set of mutually recursive fused functions, each for a different set of traversals that are executed together at some point. Furthermore each of those functions is fused independently of the others. This process introduces \emph{type-specific-partial-fusion}, since for an invocation of a traversals on a super type, the set of the called functions that corresponds to each dynamic type are fused independently, and hence some of them actually might be fusible while others are not. 
 
Note that \system only generates new functions for sequences it has not seen before. Because \system limits the number of functions that can be fused together (see Section~\ref{sec:implementation}), the number of sequences of functions is finite, and hence fusion is guaranteed to terminate.


\paragraph{Proof sketch of soundness}

The argument for the soundness of \system's fusion procedure is straightforward. First, we note that the outlining and inlining steps (steps 2 and 3) in the fusion process are trivially safe because they do not reorder any computations, and hence cannot break any dependences. Step 4 could potentially break dependences, but because \system performs a dependence analysis, it can ensure that statements are only reordered if dependences are preserved. Hence, this step is also clearly sound.

The tricky step in \system's fusion algorithm is the step where it gains the advantage of fusion: if a sequence of calls to a particular sequence of traversal functions matches a sequence that \system has already generated a fused function for, \system immediately replaces the original sequence of calls  with a call to the fused function rather than generating another new function. This is only safe if the already-fused function will do the same thing as the original function sequence. 

To see that this is safe, we note that the process of outlining followed by inlining means that the code of the fused function $f_L$ is not dependent on the node $f_L$ is invoked on---in other words, if the original sequence of traversal functions are invoked on @root.left@, after outlining and inlining, the statements within $f_L$ are relative to the formal parameter @n@ of $f_L$, and will be exactly the same as if $f_L$ were produced from the same sequence of functions invoked on a different tree node, such as @root.right@.
In other words, two identical sequences of traversal functions, $L$ and $L'$ that are invoked on different tree nodes will yield {\em identical} functions $f_L$ and $f_{L'}$ after outlining and inlining. Because the dependence graph for these functions are identical, any reordering \system does to create a fused function can be applied to both $f_L$ and $f_{L'}$. It is obvious, then, that, upon encountering the same sequence of traversal functions $L$, even if those functions are invoked on different nodes, \system can reuse an existing synthesized function.

However, there remains one gap: if, while fusing a sequence of functions $L$ to generate
$f_L$, \system encounters the same sequence of invocations $L$ that is reachable (transitively) 
from $f_L$, \system will substitute a call to $f_L$. In the simplest case,
if $f_L$ contains $L$, then a new invocation to $f_L$ will be inserted into the body 
of $f_L$. Hence, in these situations, \system is
changing the behavior of $f_L$ while using it to replace $L$.  This process feels circular. However,
a straightforward inductive argument on the depth of the call stack (in other words,
the number of recursive invocations of $f_L$ before reaching the end of the tree or some base case) 
shows that this new invocation of (the rewritten) $f_L$ behaves the same as the 
original sequence $L$. This argument mirrors the proof for the soundness of 
TreeFuser~\cite[Section 7]{sakka2017}.

\subsection{Traversal Code Generation}
\begin{figure}
\begin{grafter-numbered}

...
void _fuse__F3F4(TextBox *_r, int active_flags) {
  TextBox *_r_f0 = (TextBox *)(_r);
  TextBox *_r_f1 = (TextBox *)(_r);
  if (active_flags & 0b11) /*call*/ {
    unsigned int call_flags = 0;
    call_flags <<= 1;
    call_flags |= (0b01 & (active_flags >> 1));
    call_flags <<= 1;
    call_flags |= (0b01 & (active_flags >> 0));
    _r_f0->Next->__stub1(call_flags);
  }
  if (active_flags & 0b1) {
    _r_f0->Width = _r_f0->Text.Length;
    _r_f0->TotalWidth = _r_f0->Next->Width + _r_f0->Width;
  }
  if (active_flags & 0b10) {
    _r_f1->Height = _r_f1->Text.Length * (_r_f1->Width / CHAR_WIDTH) + 1;
    _r_f1->MaxHeight = _r_f1->Height;
    if (_r_f1->Next->Height > _r_f1->Height) {
      _r_f1->MaxHeight = _r_f1->Next->Height;
    }
  }
};
void TextBox::__stub1(int active_flags) {
    _fuse__F3F4(this, active_flags);
}
void Group::__stub1(int active_flags) {
    _fuse__F5F6(this, active_flags);
}
void End::__stub1(int active_flags) {
    _fuse__F1F2(this, active_flags);
}   
int main() {
    Group *ElementsList;
    //ElementsList->computeWidth();
    //ElementsList->computeHeight();
    ElementsList->__stub1(0b11);
}   
\end{grafter-numbered}
\vspace{-1em}
\caption{Output of \system.}
\vspace{-1em}
\label{fig:running-example-output}
\end{figure}

As described in the previous section, to fuse a sequence of functions $L$, \system 
generates a graph $G$ with statements and groups of calls that represents the fused 
function $f_L$. The body of the function $f_L$ is generated from the graph in a way similar to TreeFuser~\cite{sakka2017}, 
yet it incorporates several changes to account for mutual recursions and virtual calls.

The generated function $f_L$ is a global function that takes a pointer to the
traversed node as the first parameter (this function will be called from a 
virtual function stub placed in the tree classes). Since the  functions in $L$
can be defined in different classes, those traversals might be operating on different types,
however since they are all invoked on the same child, there must be a super type
that encloses all of them. This type is used for the traversed node parameter in
the generated function $f_L$. A lattice for the types traversed in the functions
in $L$ is created to find such type. Line 3 in figure ~\ref{fig:running-example-output}
shows the result of fusing the two functions that computes the width and the height for 
@TextBox@ element in the program shown in figure ~\ref{fig:running-example-code}.

The traversed node parameter is followed by the traversal's parameters, and an integer
that represents the set of active traversals, @active_flags@. This parameter 
can be seen as a vector of flags where each bit determines whether a specific traversal 
function is active or truncated at any point during the execution of the fused function. 
Those flags are needed because the fused traversals can have different termination conditions.

To {\em call} the the generated function $f_L$, \system replaces the original sequence 
of calls with a call to a newly created virtual function that acts as a stub and
calls the corresponding $f_L$ for each possible traversed type $T$.
Lines 26-34  in Figure~\ref{fig:running-example-output} shows an example of such stub
which is called at line 39. The stub's arguments are the arguments of the fused original call 
expression in addition to an integer that represents the active traversals. 
The stub's arguments are passed to the new fused function as well as the traversed node (\code{this}).

Statements of different traversals in the fused function should see the traversed
node type the same way they see it in the original function that they came from.
This is is needed for accessibility (a field might be defined in a derived type while
the type of the traversed node is a base type) and correctness (a derived class
can shadow a base class variable of the same name). To handle this, an alias of 
the traversed node parameter is created with the desired type for each traversal 
function using casting (lines 4  and 5 in figure \ref{fig:running-example-output})
and those aliases are used by the statements whenever a tree access happens (lines 15-19).

A topological  order of the nodes in the graph $G$ is then obtained that represents
the order of the statement in body of the fused function. Each node in the topological 
order (which corresponds to a top level statement in one of the traversals) is 
then written to the function in order. A simple statement is only executed if
the traversal that it belongs to is not terminated. Return statements terminate 
a traversal and updates the corresponding active flags as in TreeFuser~\cite{sakka2017}. 
Appropriate flags should be passed  when a fused call  is invoked within a traversal. 
The @call_flags@ variable, defined at line 7 in Figure~\ref{fig:running-example-output}, holds the active flags
that are passed to the next traversing call. Lines 7--11 fill in the appropriate
flags from the @active_flags@ in the @call_flags@ based on which traversals the outlined calls belong to.


\subsection{Limitations of \system}

Limitations of \system's dependence analyses, fusion procedure, and language mean that it cannot exploit all possible fusion opportunities for all possible traversal implementations. Here, we group the limitations of \system into three categories.

First, \system's language and implementation have been limited in some ways merely to simplify the dependence analysis. For example, \system does not support pointers other than to nodes of the tree, but relaxing this simply requires enriching the access analysis with standard alias analysis techniques. The dependence analysis can similarly be extended to support  loops within traversal functions (that do not themselves invoke additional traversal functions). In these scenarios, \system's basic fusion principles need not change. Finally, adding a shape analysis to Grafter could allow it to avoid annotating data structures to establish that they are trees.

Second, some extensions to \system would require extending the machinery of code generation to handle them. For example, supporting conditional traversal invocation can be done through syntactic manipulation (pushing the condition into an unconditionally-invoked traversal function that immediately returns if the condition is false), but this introduces instruction overhead. Managing conditional calls, traversal functions invoked within loops, or return values from traversal functions will require some new strategies for generating fused traversal code, but likely would not require substantial changes to the rest of the fusion machinery.

Finally, some extensions to \system may require devising new theories of fusion: new principles for how a fused traversal actually operates. In this category are extensions like functions that operate over multiple trees (\eg, traversals that zip together two trees), or traversals that perform more sophisticated tree mutation such as rotations.

\section{Implementation}
\label{sec:implementation}

\system is implemented as a Clang tool that performs source to source transformation for input programs
\footnote{\system is available at \url{https://bitbucket.org/plcl/grafter_pldi2019/}.}. \system uses Clang's internal AST to analyze the annotated traversals and tree structure, and performs checks to validate that the annotated  traversals satisfies \system's restrictions.
Any traversal that does not adhere to \system's language is excluded from being fused. \system uses OpenFST library~\cite{openfst} to construct the automata that represent accesses of statements and perform operations on these automata.

Different criteria can be used to perform grouping of call nodes in the dependence graph during fusion. \system uses a greedy approach for grouping: it selects an arbitrary un-grouped call node, and tries to maximize the size of the group by accumulating other un-grouped call nodes. The process continues until there is no more grouping left. This criteria is sufficient to show significant improvements (Section~\ref{sec:evaluation}), thus we did not investigate any other approach for grouping.

As mentioned earlier, in order to control the fusion process \system must limit the number of functions that can be fused together. It may seem odd that these cutoffs are needed at all, since there are only a finite number of function definitions in the program. However, if a traversal function calls {\em multiple} functions on the same child node (say, two), and each of {\em those} functions call two functions on the same child node, then it is clear that at each level of the tree, there are {\em more active traversal functions} than at the previous level. Because each step of \system's fusion process essentially descends through one level of the tree to expose more fusion opportunities, we will systematically uncover more and more functions to fuse together. Hence the need for a cutoff.
\system limits fusion in two ways: by limiting the length of a sequence of functions to fuse, and by limiting the number of times any one static function can appear in a group.


\section{Evaluation}
\label{sec:evaluation}

We evaluate \system through four case-studies from different domains, demonstrating its ability to 
express traversals over heterogeneous tree structures without compromising efficiency,
to perform fusion efficiently, and to significantly enhance the performance of traversals, 
even when processing small trees. 
The four case-studies are:
\begin{itemize}
  \item Fusing multiple traversals over a render tree.
  \item Fusing multiple AST optimization passes.
  \item Fusing multiple operations on piecewise functions.  
  \item Fusing two fast multipole method traversals.
\end{itemize}

\iftoggle{EXTND}{}
{This paper presents results for the first two case studies, which exploit the novel features of \system. The latter two case studies are Grafter versions of benchmarks from Rajbhandari et al.~\cite{rajbhandari2016a,rajbhandari2016b} and TreeFuser~\cite{treefuser}, respectively; the results are in the extended version of the paper~\cite{extended}.}

\paragraph{Experimental platform}. Since rendering is a common task performed on 
mobile phones, yet the memory on such devices is relatively small, we evaluated
the render tree traversals on a smartphone with Qualcomm Snapdragon 425 SoC.
The main platform, which is used for the other case studies, is a dual 12-core,
Intel Xeon 2.7 GHz Core with 32 KB of L1 cache, 256 KB of L2 cache, and 20 MB of L3 cache.
All cache lines are of size 64 B. 
L1, L2 caches are 8-way associative and L3 cache is 20-way associative.
We have used single-threaded execution throughout our evaluation.
Clang++ was used for compilation with "-O2" optimization level for all case studies.

For each experiment, we measure four quantities: 
\begin{enumerate}
\item The number of node visits. This measures the number of times any traversal 
function is called on any node in the tree. This provides a performance-agnostic 
measure of fusion effectiveness: the more fusion is performed, the fewer functions 
are called per tree node. 
\item The number of instructions executed. Synthesizing fused functions requires 
additional work to keep track of when various fused traversals truncate, 
additional stub virtual functions, etc. On the other hand, fusion reduces  
call and memory instructions. Measuring instructions executed provides
an estimate of \system's overall instruction overhead.
\item The number of cache misses. One benefit of fusion we expect to see is 
improved locality and reduced memory accesses. Cache misses provide a good proxy for this.

\item Overall runtime. Fusion improves locality, but potentially at the cost
 of increased instruction overhead (as reported by Sakka et al.~\cite{sakka2017}).
 The final effectiveness of fusion is hence determined by runtime.
\end{enumerate}

\subsection{Case Study 1: Render Tree}
Tree traversal is an integral part of document rendering. A render tree that 
represents the organizational structure of the document is built and traversed 
a number of times to compute visual attributes of elements of the document. 
We implemented a render tree for a document that consists of pages composed 
using nested horizontal and vertical containers with leaf elements (TextBox, Image, etc.)\iftoggle{EXTND}{}{\footnote{The extended version of the paper presents results for other types of input documents~\cite{extend}.}}.

\begin{figure}
    \centering
     \includegraphics[scale=0.24]{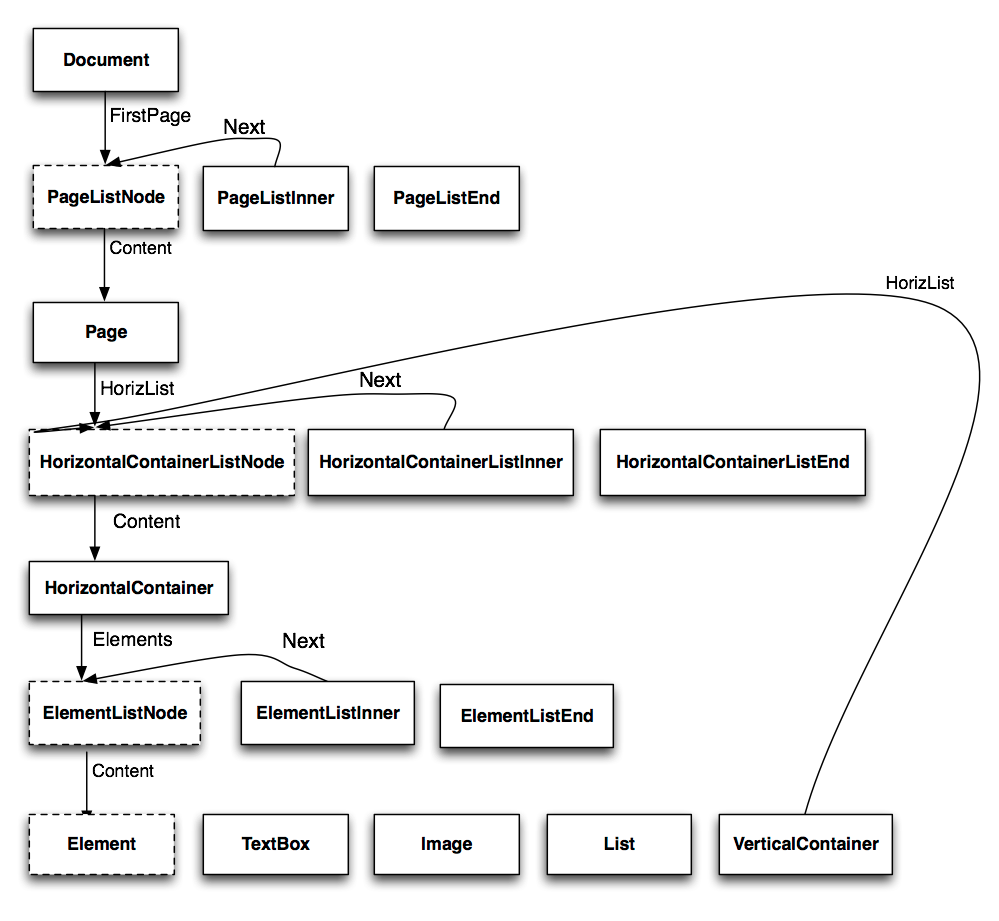}
     \vspace{-2em}
     \caption{Class hierarchy of the render tree.}
     \vspace{-1em}
     \label{fig:render-tree} 
 \end{figure}

Figure~\ref{fig:render-tree} shows the structure and the class hierarchy of the
render tree. The tree nodes are of 17 different types; boxes in the figure
represent types, arrows represent the fields and point to type of the child node. 
For instance a \code{HorizontalContainer} contains a list of elements that can 
be accessed through the field \code{Elements}. Boxes with dashed borders are
of super-types of the boxes to the right of them. For instance, an \code{Element}
can be a \code{TextBox}, a \code{Image} or a \code{VerticalContainer}. 
\iftoggle{EXTND}{
Figure~\ref{fig:page-layout} shows an example of a page that can be represented
using this class hierarchy.
}{}

\iftoggle{EXTND}{
 \begin{figure}
  \centering
   \includegraphics[scale=0.36]{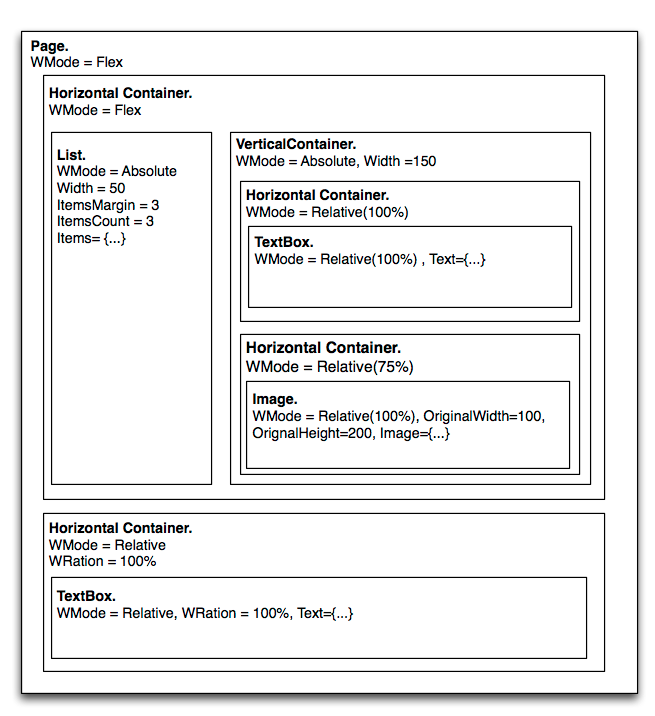}
   \vspace{-1em}
   \caption{The layout of the rendered page.}
   \vspace{-1em}
   \label{fig:page-layout}
\end{figure}
}

Five rendering passes are implemented and listed in Table~\ref{tab:passes}. 
These passes are dependent on each other. For example, computing the height of an 
element depends on computing the width and font style. In \system, passes are
implemented as fine-grained, stand-alone functions for each type. 

In the first experiment we compared the effectiveness of \system  and  TreeFuser~\cite{sakka2017},
prior work that can also perform (partial) fusion for general recursion.
We produced a baseline that performs no fusion, as well as a version that uses \system's full fusion capabilities.
We also implemented the same passes in TreeFuser~\cite{sakka2017}. 
To accommodate the limitations of TreeFuser, that implementation collapses the
types into a single type, using conditionals to determine which code path to take.
Again, we evaluate a baseline that uses the TreeFuser language to implement the passes,
and a version that performs as much fusion as possible with TreeFuser.



\begin{table}
  \begin{minipage}{0.9\linewidth}
  \centering
  \small
  \caption{Render-tree and AST passes.}
  \vspace{-1em}
  \begin{tabular}{|l|l|}
\hline
  {\bf Render-tree traversals} & {\bf AST traversals} \\
  \hline
  Resolve flex widths              &De-sugar increment \\
  Resolve relative Widths          &De-sugar decrement\\
  Set font style                   &Constant propagation     \\
  Compute height                   &Replace variable references    \\
  Compute positions                &Constant folding \\
                                   &Remove unused branches    \\
  \hline
  \end{tabular}
  \label{tab:passes}
  \end{minipage}
\end{table}

\begin{figure*}
 \begin{subfigure}[t]{0.48\textwidth}
   \includegraphics[scale=0.35]{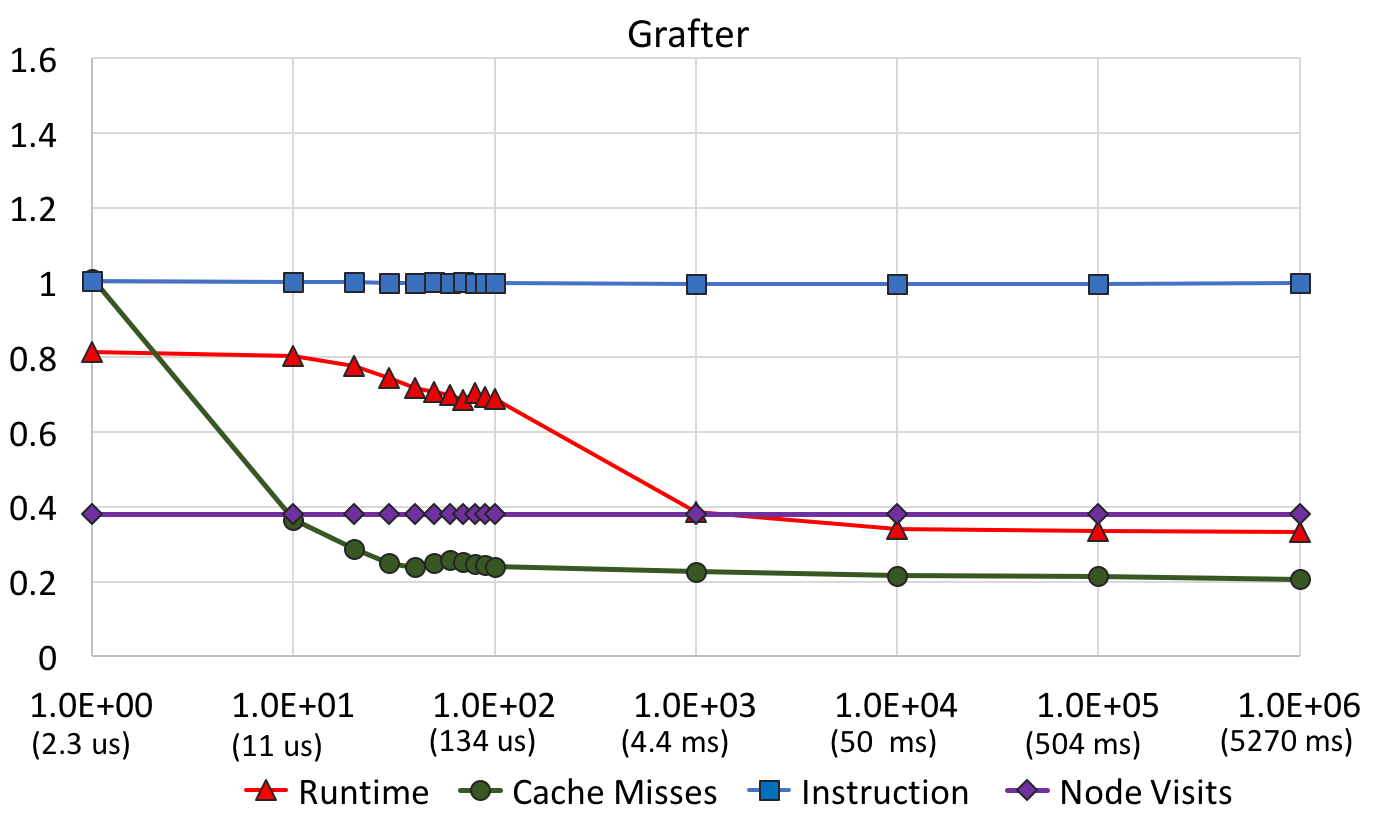}
   \caption{\system-fused implementation normalized to unfused \system implementation.}
   \label{fig:render-tree-perf-grafter}
 \end{subfigure}
 \quad
 \begin{subfigure}[t]{0.48\textwidth}
   \includegraphics[scale=0.35]{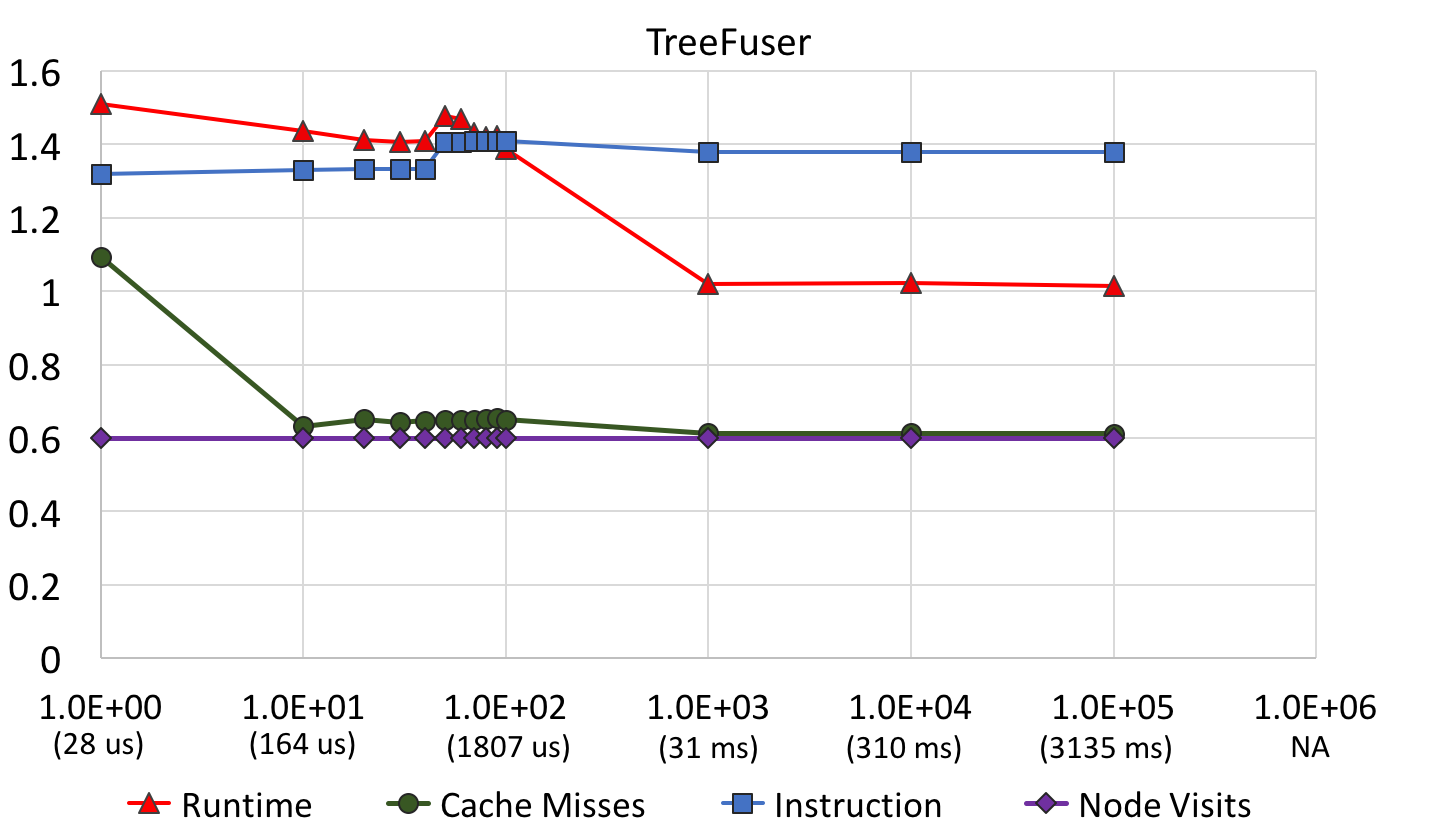}
   \caption{TreeFuser-fused implementation normalized to unfused TreeFuser implementation. (1M page baseline does not complete.)}
   \label{fig:render-tree-perf-treefuser}
 \end{subfigure} 
 \vspace{-1em}
 \caption{Performance comparison of render-tree passes written in \system and
  TreeFuser for different tree sizes. Number in parentheses is runtime of the 
  baseline. Number of pages on the x axis and 
   normalized measurement on 
  the y axis.}
   \label{fig:render-tree-perf-both}

\end{figure*}

\iftoggle{EXTND}{For this experiment we created documents of various sizes by replicating the page
shown in Figure\ref{fig:page-layout}.}{} Figure~\ref{fig:render-tree-perf-grafter} 
evaluates the \system-fused implementation of the render-tree case study, normalized
to the unfused \system baseline, while Figure~\ref{fig:render-tree-perf-treefuser} 
evaluates the TreeFuser-fused implementation, normalized to {\em its} unfused baseline%
\footnote{For small input sizes, each experiment is run in a loop to achieve a
reasonable overall runtime, then divided through by the number of loop iterations 
to find the per-run metrics. 95\% confidence intervals for render tree experiments are within 
$\pm$5\% for all measurements except for cache misses of trees smaller than 50
pages in figure \ref{fig:render-tree-perf-both}. For those the intervals expands
as the tree size decreases down to $\pm$60\%. Note that for such small trees, the absolute number of misses is quite small.}.

Both systems show the expected results of fusion: reduced node
visits after fusion, and reduced cache misses. Nevertheless, on both metrics, 
\system does better than TreeFuser: 
due to its  finer-grained representation and fusion, it can more aggressively 
fuse traversals, resulting in 60\% fewer node visits than the baseline,
compared to 40\% fewer node visits for TreeFuser. Note that because both the
\system and TreeFuser implementations do the same work, the baselines
have exactly the same absolute number of node visits. This increased fusion is reflected in cache misses: TreeFuser's fusion reduces cache misses by 40\% while \system reduces misses by 80\%. 

We also see the advantage of \system's 
type-specific fusion approach: because fused functions are on a per-type basis, 
\system is able to leverage dynamic dispatch to specialize traversals. As a result, 
it exhibits virtually no instruction overhead relative to the baseline. TreeFuser, 
in contrast, has a 30--40\% instruction overhead. 

All told, these effects mean that TreeFuser cannot achieve performance improvements
for the inputs we investigate: the improved memory system behavior cannot outweigh the instruction overhead. In contrast, \system sees substantial performance improvements:
20\% even for the smallest input (a single page), and 60\% for inputs of 1000 pages or more. 
And note that this is despite the fact that \system's {\em baseline} is already substantially 
faster than TreeFuser's, as seen in the baseline runtimes shown in Figures~\ref{fig:render-tree-perf-grafter} 
and~\ref{fig:render-tree-perf-treefuser}.

Programmability-wise, the overall logical lines of code (LLOC) 
\footnote{We use Fenton's definition of LLOC for C++: the number
of instructions with the semantic delimiter~\cite{fenton}.}
for the body of the traversals is the same for both TreeFuser and \system.
However, those LLOC are distributed among 55 different simple functions in \system while TreeFuser requires one function per traversal, with complex conditionals to disambiguate types.


\iftoggle{EXTND}{
We also evaluated \system's render tree traversal on multiple different 
documents configurations (input trees), with different properties. The results are summarized in Table~\ref{tab:render-examples}: \system achieves speedups between 1.5$\times$ and 4.5$\times$\footnote{The results in Table~\ref{tab:render-examples} were collected on the main platform, rather than the mobile platform.}.

\begin{table}
 \caption{Performance of fused traversals normalized to unfused ones for different render tree configurations. 
}
\resizebox{\linewidth}{!}{%
 \begin{tabular}{|l|r|r|r||r|l|}  \hline
 & \textbf{runtime}  & \shortstack[l]{\textbf{cache misses}}  & \shortstack[l]{\textbf{node visits}} & \shortstack[l]{\textbf{tree size}} & \textbf{description} \\
  \hline
 \textbf{Doc1}  & 0.22 & \shortstack[r]{0.17 (L2)\\ 0.14 (L3)} & 0.38 & 90MByte &\shortstack[l]{$10^5$ simple\\pages}   \\
  \hline
\textbf{Doc2}  & 0.65 & \shortstack[r]{0.28 (L2)}              & 0.4 &  4MByte & \shortstack[l]{1 dense page}\\
  \hline
\textbf{Doc3}  & 0.47 & \shortstack[r]{0.24 (L2)\\ 0.18 (L3)} & 0.4 & 58MByte &\shortstack[l]{150 pages of\\different sizes}  \\
  \hline
 \end{tabular}
}

 \label{tab:render-examples}
\end{table}
}

%
%
  
 \subsection{Case Study 2: AST Traversals}
\begin{figure}
  \centering
   \includegraphics[scale=0.30]{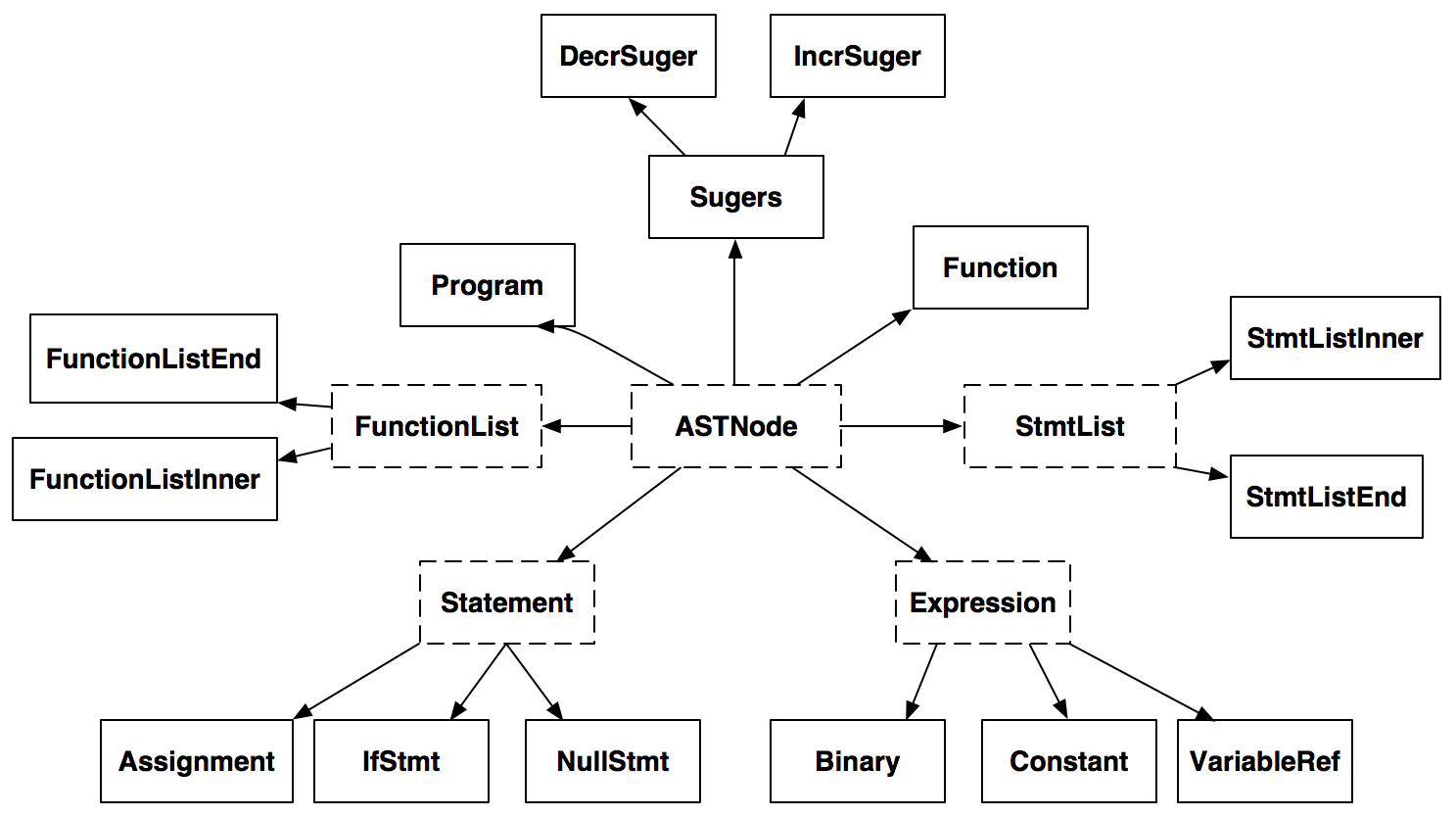}
   \vspace{-.5em}
   \caption{AST class hierarchy with 20 different types.}
   \label{fig:ast-hierarchy}
\end{figure}

Abstract Syntax Tree (AST) representation of programs is common in modern compiler frontend. 
Various validation and optimization passes are performed on AST representation.
We implemented AST passes for a simple imperative language that has assignments,
if statements, functions, and allows certain syntactic sugars. 
Figure~\ref{fig:ast-hierarchy} shows the language constructs and the class 
hierarchy of the AST that represents programs in the language. 
Node types in the AST belong to different hierarchy levels and passes are 
implemented as virtual functions defined at the top level type. 

Table~\ref{tab:passes} shows the six different AST traversal passes we 
implemented in \system: two de-sugaring passes for increment and decrement 
operations, and three optimization passes. 
The constant propagation pass is written as two traversals and works as follows: 
the constant propagation traversal looks for constant assignments and for each of 
them it initiates a traversal to replace variable references with constants.
The AST passes depends on each other; de-sugaring must happen before the optimization
passes, and removing unused branches depends on constant folding and propagation 
since they may produce constant branch conditions. 
Furthermore, those passes mutate the tree (e.g., to de-sugar an expression, one part
of the AST is deleted and another part is constructed).

\begin{figure} 

  \centering
   \includegraphics[scale=0.37]{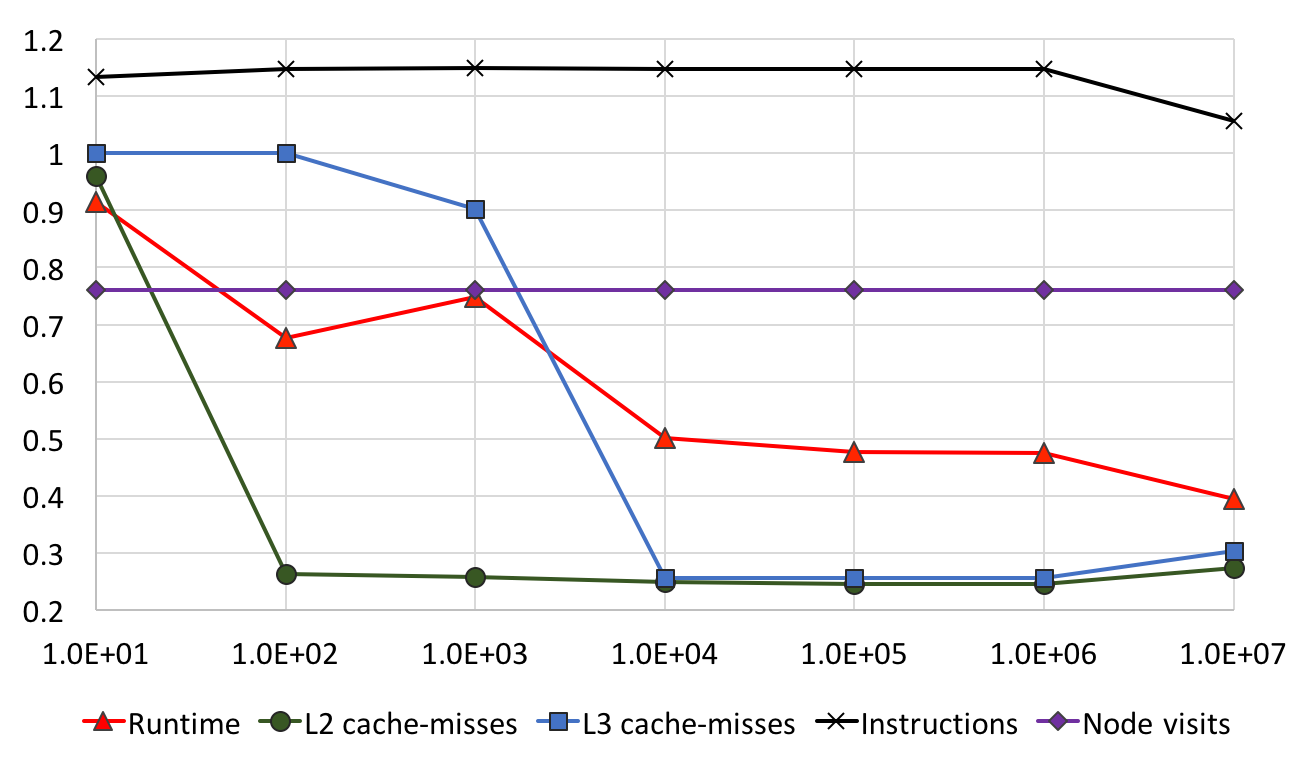}
   \vspace{-1em}
   \caption{Performance measurements for fused AST traversals normalized to the unfused ones for different tree sizes. Number of functions on the $x$ axis, and normalized measurements on the $y$ axis.}
   \label{fig:ast-perf}
\end{figure}

We wrote a function that consists of different statement types, and expressed it as an AST.
This function was replicated in order to obtain bigger trees for the evaluation\iftoggle{EXTND}{}{\footnote{The extended version of the paper presents additional results for different input configurations~\cite{extend}.}}.
Figure \ref{fig:ast-perf} shows the performance of the fused AST traversals with
respect to the un-fused ones \footnote{95\% confidence intervals for the AST experiments are within $\pm$5\%
for all measurements, except for the point $10^3$ in Figure~\ref{fig:ast-perf}, where confidence intervals 
are within $\pm$33\% for L3 cache misses and $\pm$15\% for runtime.}. 
\system reduces L2 cache misses by 75\% and once the tree is big enough, it reduces L3 misses by 70\% as well. 
The fused traversals have  instruction overhead (between 15\% and 4\%), but that overhead is overcome by the reduction in cache misses. 
The fused traversals are 1.25$\times$ to 2.5$\times$ faster than the un-fused traversals depending on the tree size.

The instruction overhead is caused by different AST passes having  different truncation conditions, unlike the render tree traversals, which all completely traverse the tree and get truncated at the same time. 
For instance, the ``replace variable references'' pass gets truncated once the reference is reassigned. Because this truncation is dynamic,
parameters of the truncated traversals will kept being passed, and the truncation flags will continue to be checked, until all traversals truncate, increasing overhead.

\iftoggle{EXTND}{
Table \ref{tab:ast-examples} shows the performance of the fused traversals for
different AST inputs with respect to the unfused ones. \emph{Prog1} consists of a large number of normal-sized 
functions. Since all the traversals are fusible across the function list, \emph{Prog1} 
has the highest reduction in node visits, 34\%. On the other hand \emph{Prog2} consists 
of only one large function, and hence has less reduction in node visits, 8\%.
\emph{Prog3} consists of multiple functions with long live ranges. The tree of \emph{Prog3} is the 
largest and thus reductions in L3 cache misses are achieved along with a 70\% reduction in the runtime.  

\begin{table}
  \caption{Performance of fused traversals normalized to unfused ones for different AST configurations}
\resizebox{\linewidth}{!}{
 \begin{tabular}{|l|r|r|r||r|l|}  \hline
 & \textbf{Runtime}  & \shortstack[l]{\textbf{Cache misses}}  & \shortstack[l]{\textbf{Node visits}} & \shortstack[l]{\textbf{Tree size}} & \textbf{Description} \\
  \hline
 \textbf{Prog1}  & 0.71 & 0.26 (L2) & 0.76 & 8MByte &Small functions \\
  \hline
\textbf{Prog2}  & 0.87 & 0.58 (L2) & 0.92 & 6MByte & One large function\\
  \hline
\textbf{Prog3}  & 0.31 & \shortstack[r]{0.48 (L2)\\ 0.26 (L3)} & 0.93 & 31MByte &\shortstack[l]{Long live ranges}  \\
  \hline
 \end{tabular}
}

  \label{tab:ast-examples}
\end{table}
}
{}

\iftoggle{EXTND}
{
\subsection{Case Study 3: Piecewise Functions}
\emph{MADNESS (Multiresolution, Adaptive Numerical Environment for Scientific Simulation)}~\cite{madness} uses kd-trees to compactly represent piecewise functions over a multi-dimensional domain.
The inner nodes of the tree divide the domain of the function into different sub-domains,
while leaf nodes store the coefficients of a polynomial that estimates the function 
within the node's sub-domain. MADNESS uses this representation to solve differential and integral equations: mathematical operations on these functions (\eg, differentiation, scaling, sqaring) can be captured as traversals of the tree.

In this case study, we  implemented kd-trees for single variable functions, 
and different traversals to perform computations on these functions. 
Table~\ref{tab:piecewise-operations} shows these traversals. 
Some of these traversals requires structural changes to the tree. 
For example, the \emph{multXRange($a$, $b$)} traversal multiplies the function in the range $[a, b]$ by $x$ where $x$ is the variable representing the domain. 
A leaf node that has a non-empty intersection interval of its domain and the interval $[a, b]$, 
needs to be split into multiple nodes if its domain does not lie completely  within the interval $[a, b]$.

\begin{table}[]
  \caption{Traversals over piecewise functions.}
\resizebox{0.7\linewidth}{!}{
 \begin{tabular}{|l|l|} \hline
\textbf{Function} & \textbf{Description} \\
\hline
scale($c$) & $f(x) = cf(x)$ \\
\hline
add($c$)   & $f(x) = f(x)+c$ \\
\hline
square() &$f(x)=f(x) \cdot f(x)$\\
\hline
differentiate()& $f(x) = f^{(1)}(x)$ \\
\hline
addRange($c$, $a$, $b$) & $f(x)=f(x)+c(u(a)-u(b))$ \\
\hline
\shortstack[l]{multXRange($a$, $b$)\\ \ } & \shortstack[l]{$f(x)=xf(x)\cdot(u(a)-u(b))$ \\
                                                                 $+f(x)\cdot(1-u(a)+u(b))$ } \\
\hline
addXRange($a$, $b$) & $f(x) =f(x)+x(u(a)-u(b))$\\
\hline
integrate($a$, $b$) & $\int_{a}^{b} f(x)$ \\
\hline
project($x_{0}$) & $f(x_{0})$\\
\hline
\end{tabular}
}
\label{tab:piecewise-operations}
\end{table}

Unlike the previous case studies, the schedule of traversals in this case-study depends on the constructed equation and differs from one another. 
Hence, manual fusion of such traversals is not practical, because it needs to be done for each equation separately based on the different operations in the equation. 
We constructed three different equations that use different schedules of traversals, shown in Table~\ref{tab:piecewise-perfom}.
We evaluated the performance of the fused traversals for each of these equations on a balanced kd-tree constructed by uniformly partitioning the interval [$10^5$,$10^{-5}$].

Figure~\ref{fig:kd-perf} shows the performance of fused traversals corresponding to the first equation in Table~\ref{tab:piecewise-perfom} for different depths of kd-tree, normalized to the unfused ones.
\footnote{95\% confidence intervals for kd-tree experiments are within $\pm$1\% for all reported measurements except for L3 cache misses of trees of depth 16, reported in Figure~\ref{fig:kd-perf}, where intervals are $\pm$16\%.}
The fused traversals reduce node visits by 83\%, and we see a 90\% reduction in L2 cache misses. 
Overall, the fused traversals are faster, with a runtime improvement ranging from 15\% for small trees, to 66\% for large ones.
Table~\ref{tab:piecewise-perfom} summarizes the performance of the corresponding fused traversals for each equation normalized to the unfused ones when performed on a balanced kd-tree of depth 20.

\begin{figure}
  \centering
   \includegraphics[scale=0.33]{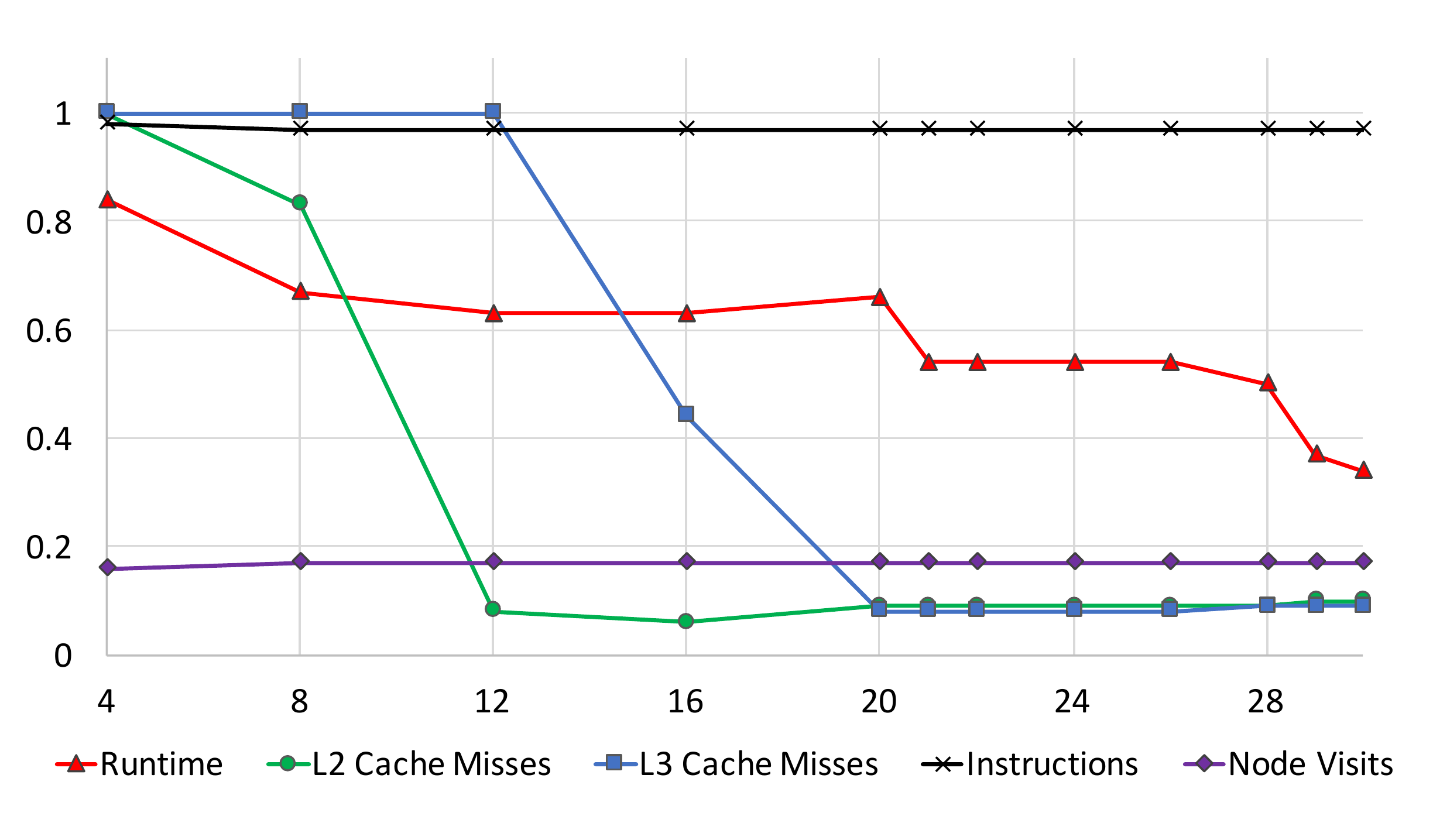}
   \vspace{-1em}
   \caption{Performance measurements for fused kd-tree traversals normalized to the unfused ones for different tree sizes. Depth of the tree on the $x$ axis, and normalized measurement on the $y$ axis.}
   \label{fig:kd-perf}
\end{figure}

\begin{table}
 \caption{Performance improvement of different fused traversals on piecewise functions.}
\resizebox{\linewidth}{!}{
\begin{tabular}{|l|r|r|r|}  \hline
 \textbf{Equation} &\textbf{Runtime}  & \shortstack[l]{\textbf{Cache misses}}  & \shortstack[l]{\textbf{Node visits}}  \\
 \hline
 $x^4(f^{(2)}(x))^2+\sum_{i=0}^{3}x^i$ & 0.66& 0.09(L2) 0.51(L3)  & 0.17 \\
  \hline
 $f^{(5)}(x)|_{x=0}$ & 0.49& 0.20(L2) 0.51(L3) &0.20\\
  \hline
 $\int_{-10^5}^{10^5} x^3(f(x)+.5)^2\cdot u(0)$ & 0.88 & 0.33(L2) 0.65(L3) &0.33\\
  \hline
\end{tabular}
}
 
  \vspace{-1em}
\label{tab:piecewise-perfom}
\end{table}
}{}

\iftoggle{EXTND}
{
\subsection{Case Study 4: Fast Multipole Method}
The fast multipole method (FMM) is a numerical technique used in evaluating pairwise interactions
between large number of points distributed in a space (\eg~Long-ranged forces in the n-body problem, computation of gravitational potential, and computational electromagnetic problems)~\cite{greengard87}.

In this case study, we reimplement the FMM benchmark from TreeFuser in \system, which is based on the implementation from the Treelogy benchmark suite~\cite{hegde17ispass}.
Figure~\ref{fig:fmm-perf} shows the performance of fused traversals for different input sizes.
Grafter was able to fully fuse the two passes and yield a performance improvement up to
22\% over the unfused version.\footnote{95\% confidence intervals for the FMM experiments are within 
$\pm$1\% for all measurements.}


\begin{figure}
  \centering
   \includegraphics[scale=0.4]{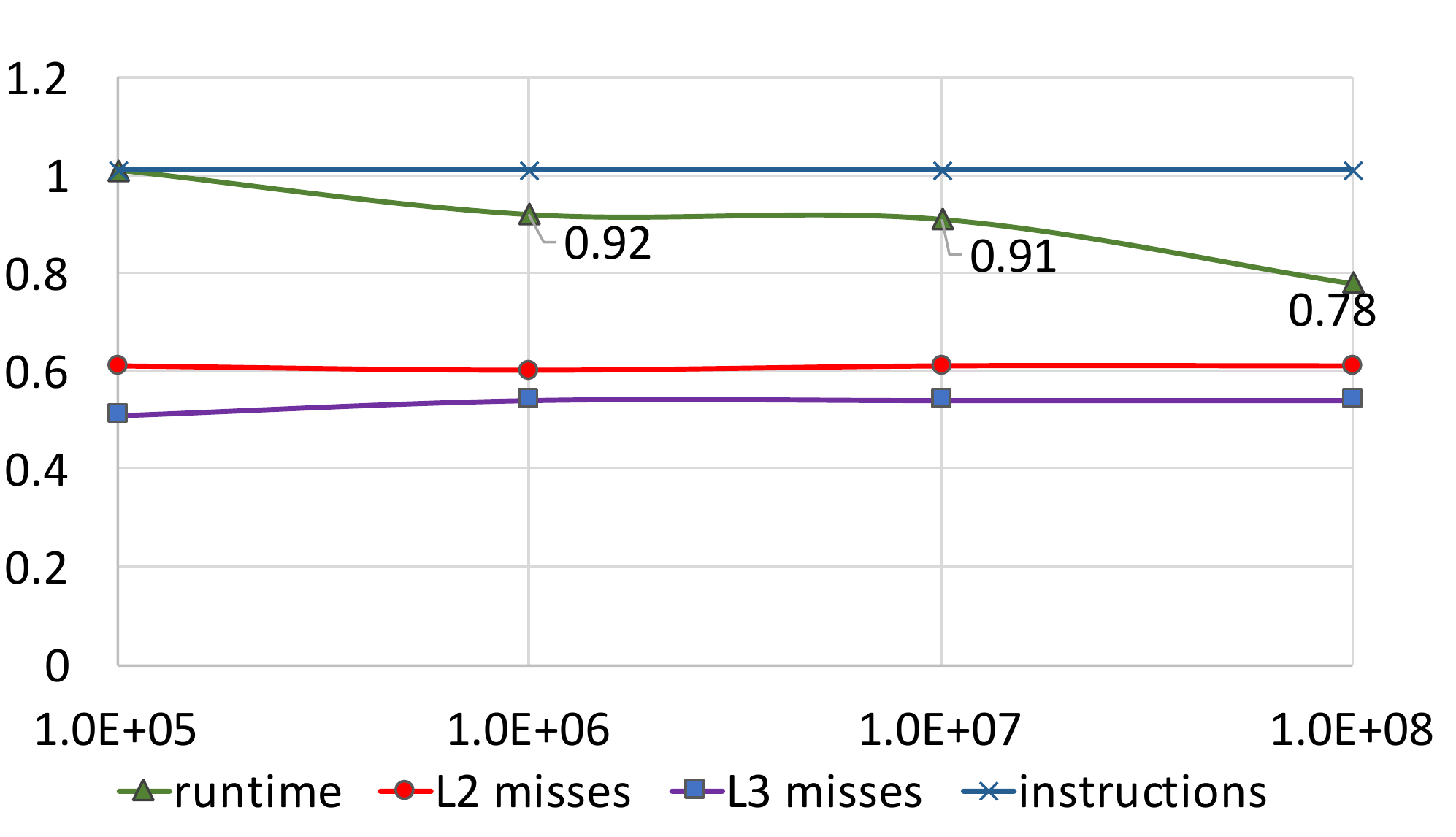}
   \vspace{-1em}
   \caption{Performance measurements for fused FMM traversals normalized to the unfused ones for different number of points. Number of points on the $x$ axis and normalized measurement on the $y$ axis.}
   \vspace{-1em}
   \label{fig:fmm-perf}
\end{figure}
}{}

\section{Related Work} \label{sec:related}

The overall goal of this work is to enable programmers to achieve efficient
tree traversals.  Thus it is obliquely related to prior work on that improves
the performance of a \emph{single} tree traversal by adjusting memory layout or
encoding \cite{chilimbi1999,Truong1998,Lattner2005,Chilimbi1999a}.
%
The most directly related work, however, optimizes series of traversals,
typically by fusing together multiple passes.  This work can be sub-divided
based on the \emph{source representation} it targets:


\paragraph{Specialized or domain-specific programs}
In these systems, the programmer aids the tool by expressing their tree
traversals in a structured form.  For example, in {\em miniphases}
\cite{petrashko2017}, the programmer writes an AST transformation as a
collection of ``hooks'' or callbacks that observe individual syntax nodes and
act on them, while abstracting the recursion over the tree so that it is
handled by the underlying framework.
The user manually groups these miniphases into tree traversals, but there is no
guarantee of equivalence between fused and unfused executions---that soundness
burden is on the programmer.
This style of decomposing passes is often approximated in many compiler
implementations---strategies for it are part of compiler writer folklore.  It
also appears explicitly in software frameworks such as the a {\em nanopass
  framework}~\cite{nanopass}.

Another specialized way to express tree traversals is by using attribute
grammars to specify {\em tree transducers}---automata that traverse trees and
create output~\cite{tata2007,doner1970}.
Using this specialized representation of programs, subsequent authors were able
to improve efficiency of compositions of tree
transducers~\cite{engelfriet02,maletti08}.
However, to our knowledge none of these approaches handle as general a class of
programs and fusion opportunities as \system: for example, we are not aware of
a tree transducer solution for partial fusion.

Enabling fusion based on a high-level representation of traversals has likewise been
explored in the HPC domain: for instance the MADNESS simulation framework uses
this approach for fusing kd-tree passes
\citet{rajbhandari2016b,rajbhandari2016a}.  This abstraction supports a
more restrictive notion of fusion than \system. All traversals must visit the
same children and must visit them in either a pre-order or post-order; partial
fusion is not possible. Moreover, the framework relies on programmer-provided dependence information.

\paragraph{Functional Programs}
As discussed in the introduction, the prior research explores fusing general recursive
functional programs and eliminating intermediate data structures (deforestation)~\cite{wadler1990}.
In practice, functional programming languages have settled on using libraries of
combinators with known fusion transformations (like @map@ and @fold@).  For
example, in the Haskell ecosystem, many everyday libraries use the {\em stream
  fusion} approach~\cite{stream-fusion} and variations on it.
This combinator style works well for collections (arrays, lists, etc) but is
sharply more limited than general recursive functions on trees. One example of this
fact is that compiler writers cannot use such frameworks for fusing AST traversals.

\paragraph{Imperative Programs}
The most closely related works to \system are systems that target imperative or
object-oriented source programs.
Aside from TreeFuser (which we have already described) most work on fusion in
imperative programs focuses on {\em loop fusion}, merging the bodies of adjacent
loops.  Typically this is applied to programs operating on arrays and matrices
\cite{darte2000complexity,kennedy1993maximizing,qasem2006profitable}.  The most
well known work in this area targets the further restricted class of programs
with {\em affine} indexing expressions~\cite{bondhugula08}.
These approaches do not generalize to recursive programs operating over irregular
data such as the tree traversals we explore in this paper.

\section{Conclusions}
\label{sec:conclusions}

This paper introduced \system, a framework for performing fine-grained fusion of generic tree traversals. In comparison with prior work, \system is either more general (in its ability to handle general recursion and heterogeneity), more effective (in its ability to perform more aggressive fusion), sound (i.e., without relying on programmer assumptions of fusion safety), or some combination of all three. We showed that \system is able to effectively fuse together traversals from two domains that rely on repeated tree traversals: rendering passes for document layout, and AST traversals in compilers. Not only can \system perform more aggressive fusion than prior work, it also delivers substantially better performance. \system allows programmers to write simple, ergonomic tree traversals while relying on automation to produce high-performance fused implementations.

\begin{acks}                            
  This work was supported in part by \grantsponsor{nsf}{National Science Foundation}{} awards \grantnum{spx-milind}{CCF-1725672}{} and \grantnum{spx-ryan}{CCF-1725679}, and by \grantsponsor{doe}{Department of Energy}{} award \grantnum{doe-career}{DE-SC0010295}. We would like to thank our shepherd, Ben Titzer, for his help with the revisions of this paper, as well as the anonymous reviewers for their suggestions and comments.
\end{acks}

\bibliography{paper}

\end{document}